\documentclass[referee]{aa}  

\usepackage{graphicx}

\usepackage{natbib}
\usepackage{aas_macros}
 
\def\e20{$\times 10^{20}$}

\def\today{\number\day -\number\month -\number\year}
\def\MCG{MCG-05-07-001 }
\def\MCGa{MCG-05-07-001}
\def\n12{NGC~1210}

\def\bigskip{\vskip 12pt}

\def\smallskip{\vskip 8pt}

\begin{document}

\title{{\it GALEX} UV properties of the polar ring galaxy \MCG \\ and  the
shell galaxies NGC~1210 and NGC~5329
\thanks{Based on  {\it GALEX}  observations: GI1-059 PI D. Bettoni}}

\author{
 A. Marino\inst{1,5}, E. Iodice\inst{2},  R. Tantalo\inst{3},  L. Piovan\inst{3}, \\
 D. Bettoni\inst{1}, L.M. Buson\inst{1}, C. Chiosi\inst{3}, G. Galletta\inst{3},
 R. Rampazzo\inst{1} \and R.M. Rich\inst{4}
      }

\offprints{amarino@pha.jhu.edu}

\institute{
          INAF-Osservatorio Astronomico di Padova,
          Vicolo dell'Osservatorio 5, I-35122, Padova, Italy \\
          \email{antonietta.marino@oapd.inaf.it; daniela.bettoni@oapd.inaf.it;\\
          lucio.buson@oapd.inaf.it; roberto.rampazzo@oapd.inaf.it}
     \and
          INAF-Osservatorio Astronomico di Capodimonte,
          via Moiariello 16, 80131, Napoli, Italy \\
          \email{iodice@na.astro.it}
     \and
          Dipartimento di Astronomia Universit\`a di Padova,
          Vicolo dell'Osservatorio 2, 35122 Padova, Italy \\
          \email{cesare.chiosi@unipd.it; giuseppe.galletta@unipd.it; \\ rosaria.tantalo@unipd.it, lorenzo.piovan@gmail.com}
     \and
            Physics \& Astron. Dept., UCLA,
            Box 951562, 405 Hilgard Ave., Los Angeles CA 90025-1562, USA \\
             \email{rmr@astro.ucla.edu}
     \and
         Department of Physics and Astrononomy, Johns Hopkins University
            3400 North Charles Street, Baltimore, MD 21218, USA \\
             \email{amarino@pha.jhu.edu}
            }

\date{Received / accepted}

\abstract
{Systems of shells and polar rings in early-type galaxies are
 considered ``bona fide" tracers  of mass accretion and/or mergers.
Their high frequency in low density environments suggests that these
processes could drive the  evolution of at least a fraction of the
early-type galaxy population.}
{ We investigate the star formation histories of this type of galaxies.
 Their UV emission is important for testing whether these
galaxies host ongoing  or recent star formation  and  how this formation varies
across the galaxy.}
{We used far- and near- ultraviolet, optical, near-infrared images,
neutral hydrogen HI maps, and line-strength indices to investigate
the nuclear and outer regions of these galaxies as well as  the
regions where fine structures are present.}
{ The GALEX Near UV (NUV) and Far UV (FUV) images of  \MCG and
NGC~1210 show complex tidal tails and debris structures. The far UV
morphology of both galaxies appears so different from the optical
morphology that  the early-type classification may not apply.    
In both GALEX bands, the polar ring of \MCG is the dominant feature, whereas
an extended tidal tail dominates the FUV bands of NGC~1210. In \MCG
and NGC~1210, there is a strong correlation between structures
detected in the FUV and NUV bands and in HI. In contrast, NGC~5329
does not show evidence of shells in the GALEX bands.  We try  to
constrain the age of the accretion episode or merger that produced to the
shells and polar rings with the aid of composite stellar populations
that take the presence of dust into account. The presence of HI in
both \MCG and NGC~1210 argues  in favour of wet mergers.
Models suggest the presence of very young stellar populations in
\MCGa: the observations could be explained in the framework
of a conspicuous burst of star formation that occurred $\leq$1 Gyr
ago and involved a large fraction of the galaxy mass. Our models
suggest that also the nuclei of NGC~1210 and NGC~5329 could have
been  rejuvenated by an accretion episode about 2-4 Gyr ago. }

{}

  \keywords{Ultraviolet: galaxies  -- Galaxies: elliptical, lenticular, CD
-- Galaxies: individual: \MCGa, NGC~1210, NGC~5329   -- Galaxies: evolution}

\titlerunning{{\it GALEX} UV of \MCGa, NGC~1210, and NGC 5329}
\authorrunning{Marino et al.}
\maketitle

\section{Introduction}\label{intro}

Both early-type galaxies hosting polar rings and those showing {\it
fine structure}, such as shells, are considered the site of (sometimes
major)  mass accretion events. In the hierarchical evolutionary
scenario they play a special role: namely  to fill the gap between
ongoing mergers and the quiescent elliptical galaxies.

Polar ring galaxies (PRGs) are peculiar objects consisting of a
 central spheroidal component, the {\it host galaxy (HG)}, surrounded
by an outer {\it ring (PR)}, made up of  gas, stars, and dust, that
the central galaxy \citep{Whitmore90}. PRGs are quite rare: they
amount to only 0.5\% of early-type galaxies in the Local Universe. Being
quite difficult to detect, \citet{Whitmore90} estimated that 4.5\% of
the nearby lenticular galaxy population have  polar rings. They also
suggested that the fraction of galaxies that possessed a polar ring at
some point in their history could be higher. Indeed, some galaxies
may have lost their polar ring because the stability of the polar
structure depends also on the impact parameters \citep{Katz92,
Christodoulou92}.  Until some years ago, the most credited formation
scenarios for PRGs were either  gas accretion by infall or the merger of
two gas-rich disk galaxies \citep{Thakar96, Thakar98, Reshetnikov97,
Bekki98, Bournaud03, Bournaud05}. In the accretion scenario, the
accreted material will form a ring that settles into one of the
principal planes of the gravitational potential of the host galaxy
\citep{Bournaud03}.  The merger of two unequal-mass disk galaxies is
another way of forming a PRG: the ``intruder", on a polar orbit with
respect to the ``victims" disk, passes close to its center, decelerating 
and pulling itself back toward the victim by strong
dissipation caused by the interaction with the ``victim"
gaseous disk. The morphology of the merger remnants depends on the
merging initial orbital parameter and the initial mass ratio of the
two galaxies \citep{Bekki98, Bournaud03}. Theoretical studies
found that an alternative way of producing a PRG is by means of external gas
accretion from the cosmic web filaments with inclined angular
momentum \citep{Dave01, Semelin05, Maccio06}.  
By using high-resolution cosmological simulations of galaxy formation in the
standard CDM scenario,  it has been shown that  angular
momentum misalignment during  hierarchical structure formation
can lead to the formation of a high inclined ring/disk
\citep{Brook08}.

\begin{table*}
\caption{Relevant photometric, structural, and kinematic  properties of \MCGa, NGC 1210, and NGC 5329 }
\label{table1}
\center
\begin{tabular}{lcccc}
\hline
&&&&\\
                                          & \MCG          &   NGC 1210      & NGC5329        &  Ref. \\
&&&&\\
\hline
Other identifications                               & ESO~415~G26          &  ESO 480- G 031 & WBL~472-001    &  \\
                                          & AM 0226-320          &  AM 0304-255    &                &       \\
\hline
Morphological Type                              &     S0-pec           &(R')SB(rs)- pec  & E              & [1]   \\
Mean Hel. Sys. Vel. [km~s$^{-1}$]          &  4604$\pm$14         &  3878$\pm$7     & 7109$\pm$31    & [1]   \\
Adopted distance [Mpc]                     &   61.4               &   51.7          &  94.8          &       \\
B$_T$                                      &  14.70$\pm$0.13      & 13.46$\pm$0.19  & 13.37$\pm$0.15 & [1]   \\
                                          &                      &                 &                &       \\
{\bf Galaxy structure}:             &                      &                 &                &       \\
Average ellipticity                        &      0.65            &    0.59         &  0.48          & [2]   \\
P.A.  [deg] main body                      &        22            &    117.7        &                & [2]   \\
P.A. (ring)  [deg]                         &      94.2            &                 &                & [3]   \\
HI total mass [M$_\odot$]                  & 5.6 $\times $ 10$^9$ &                 &                & [4]   \\
H$_2$ total mass [M$_\odot$]               & 2.4 $\times $ 10$^9$ &                 &                & [5]   \\
                                          &                      &                 &                &       \\
{\bf Kinematic parameters}        &                      &                 &                &       \\
Vel.disp. $\sigma_0$ stars [km~s$^{-1}$]   &   127$\pm$3          & 219$\pm$27[6]   &262$\pm$22      & [2]   \\
Max. rotation V$_{max}$ star [km~s$^{-1}$] &   145$\pm$12         & $>$129[6]       &                & [2]   \\
\hline
\end{tabular}

\medskip

References: [1] {\tt NED http://nedwww.ipac.caltech.edu/};
[2] {\tt HYPERLEDA http://leda.univ-lyon1.fr/};
[3]  \citet{vanGorkom87};
[4]  \citet{Schiminovich97};
[5]  \citet{Galletta97};
[6]  \citet{Longhetti98a}.
\end{table*}

Among {\it fine structures}, shells are faint, sharp-edged stellar
features \citep{Malin83} characterizing a significant fraction
($\approx$ 16.5\%) of the field early--type galaxies
\citep{Schweizer93a, Reduzzi96, Colbert01}.  Different scenarios for
their origin emerge from the rich harvest of simulations performed
since their discovery in the early 80's: from major mergers
\citep{Barnes92, Hernquist92, Hernquist95}, to accretion between
galaxies of different masses (mass ratios typically 1/10 - 1/100)
\citep[see e.g.][]{Dupraz86, Hernquist87a, Hernquist87b}, and  to even
weaker interaction events \citep{Thomson90, Thomson91}.  A few
models invoke internal mechanisms for  shell formation such us gas
ejection  powered by either the central AGN or supernova explosions
\citep{Fabian80, Williams85}. In the accretion model, shells are
density waves formed by the infall of stars from a companion. The
accretion/merger mechanisms qualitatively reproduce the basic
characteristics, such as spatial distribution, frequency, and shape
of observed shell systems \citep[see e.g.,][and references
therein]{Wilkinson00, Pierfederici04, Sikkema07}.  The inner
mechanisms cannot correlate  the shell formation with the environment,
whereas the accretion/merger models can explain why shells are
predominantly found in low density environments
\citep[see][]{Malin83}. Low-density environments can indeed favour the
occurence of merger/accretion events, because the group velocity dispersion is of
the order of the internal velocity of the member galaxies
\citep{Aarseth80, Barnes85, Merritt85}.

Observations in the past two decades have proven that a significant
fraction of early-type galaxies contain gas in the ionized, atomic,
and molecular phases. The study of the cold and warm gas phases  provides 
new insight into the evolution of shell galaxies.
\citet{Rampazzo03} found that the warm ionized (H$\alpha$) gas and
stars  often appear dynamically decoupled indicative of external
acquisition of the gas, as predicted by merger mechanisms
\citep{Weil93}. In the shell galaxy IC~4200, \citet{Serra07a} showed that 
the ionized gas is decoupled from stars and  its
rotation might be a continuation of the HI velocity field. In contrast, in other shell
galaxies, e.g., NGC~2865, a clear association between the cold
(HI/CO) gas and  the kinematics of stars is evident, which appears to be 
inconsistent with the merger picture \citep{Schiminovich94, Schiminovich95,
Charmandaris00, Balcells01}.

The pioneering study of \citet{Schweizer92} showed that, on 
average, field early-type galaxies with fine structures are
significantly younger ($\sim 4.6$ Gyr) than those without ($\sim 8$
Gyr). The presence of fresh gas could lead to ``rejuvenations" of the
stellar populations,  which may  be used to estimate the time
and duration of the polar ring and shell phenomena. The star
formation history of shell early-type galaxies was analysed by
\citet{Longhetti00} using  line--strength indices.  They showed
that shell-galaxies have a wide range of ages inferred from
the H$\beta$ versus MgFe plane, which implies that recent and
old interaction/acquisition events are equally probable.  If shells
formed at the same time as the ``rejuvenating" event took
place, shells ought to be long--lasting phenomena.  
Combining {\it GALEX} far UV data and
line--strength indices, \citet{Rampazzo07} showed that the peculiar 
position of some shell galaxies in the (FUV-NUV) versus  H$\beta$ plane 
could be explained in terms of a recent (1-2 Gyr old) rejuvenation episode. 
Based on the line-strength indices, \citet{Serra07a} suggest that the origin of
IC~4200, a HI-rich shell galaxy, was a major merger event that occurred
1-3 Gyr ago.

In this framework, we investigate the ability of {\it GALEX}
(FUV-NUV) colours to expand our view of stellar populations  
in galaxies that are sites of accretion/merger events such us the polar
ring and shell galaxies. More specifically, for ring galaxies we
 estimate the age of the ring from the mix of stellar
populations therein,  while for  shell galaxies  we attempt to gain 
information about the ``rejuvenation" process.

We present a study of shell galaxies that completes our Cycle~1
set of {\it GALEX} observations (ID=GI1-059). The UV observations
include  the polar ring galaxy \MCG and NGC~1210.   We also include
NGC~5329,  which  is  in the list of galaxies with shells  
in the survey of \citet{Malin83}. These authors do not provide a description of
the shell system, in contrast to  most other cases in the catalogue,
but note that NGC~5329 is located in a group ($\approx$ 10
galaxies) environment similar to that of both NGC~2865 and  
NGC~5018 studied by \citet{Rampazzo07}. 
This paper  extends the UV study by \citet{Rampazzo07}. For this purpose, 
(1) we  double the sample size of shell galaxies; (2) we consider shell 
galaxies with cold gas at their centre, enlarging the range of physical properties
investigated; (3) we use detailed chemo-spectro-photometric models to study the 
colours of \MCG derived from the observations that extend from the far UV to 
near infrared.

The plan of the paper is as follows. Section ~\ref{sample} presents
the sample  and summarizes the current
literature on the subject. Section~\ref{observ} presents 
the observations. The main results of the
observations are presented in Sect.~\ref{results}.  An attempt to
interpret the results with particular emphasis on the UV colours
and their significance is given in Sect.~\ref{theo_vs_obs}.
Finally, our conclusions are summarized in  Sect.~\ref{sum_concl}.
 We assume that H$_0$=75 km/s/Mpc  throughout the paper.

\begin{figure*}
\center
{\includegraphics[width=14.9cm]{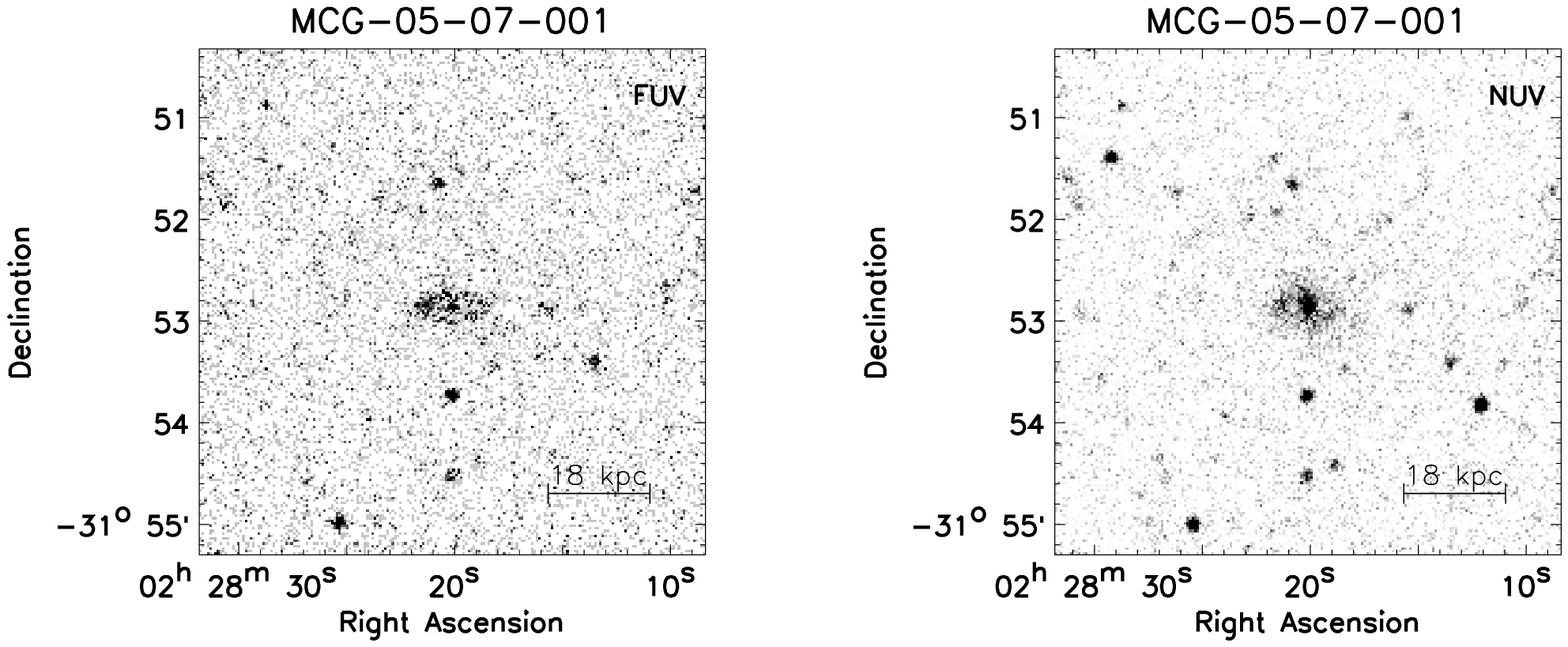}}
{\includegraphics[width=14.9cm]{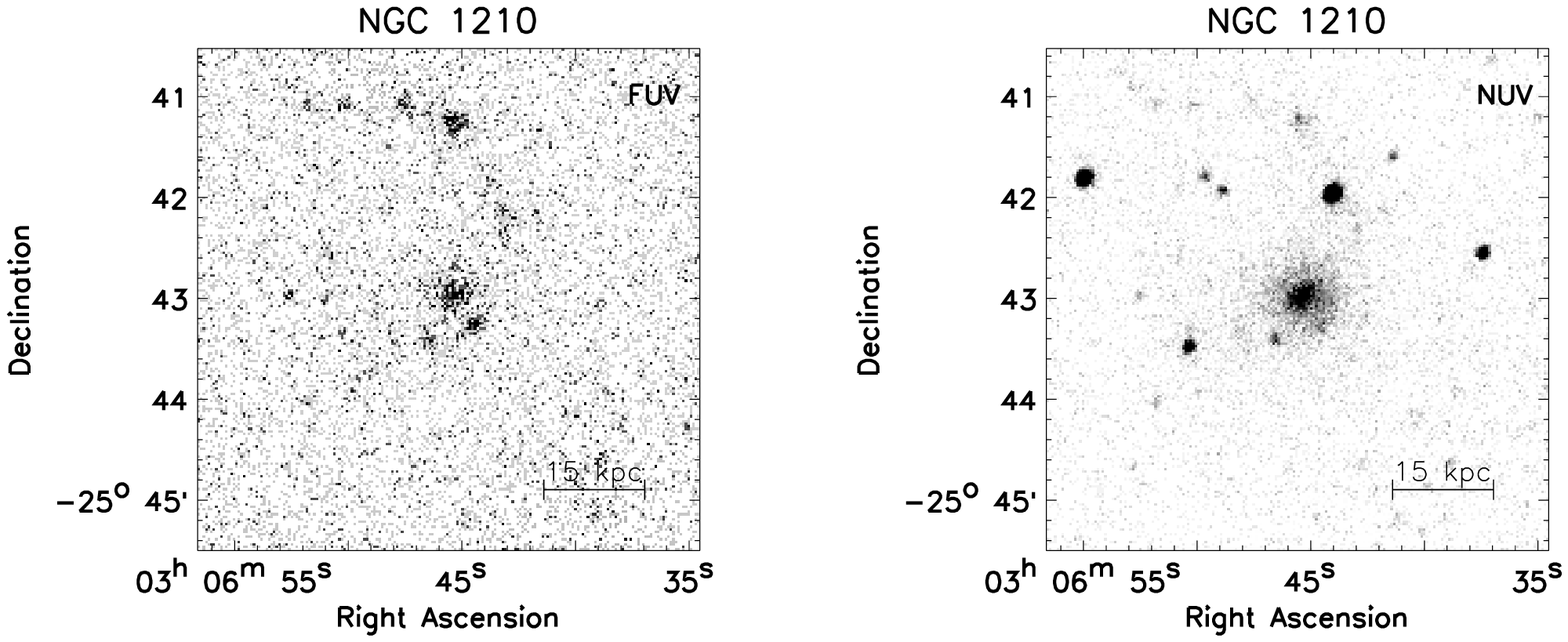}}
{\includegraphics[width=14.9cm]{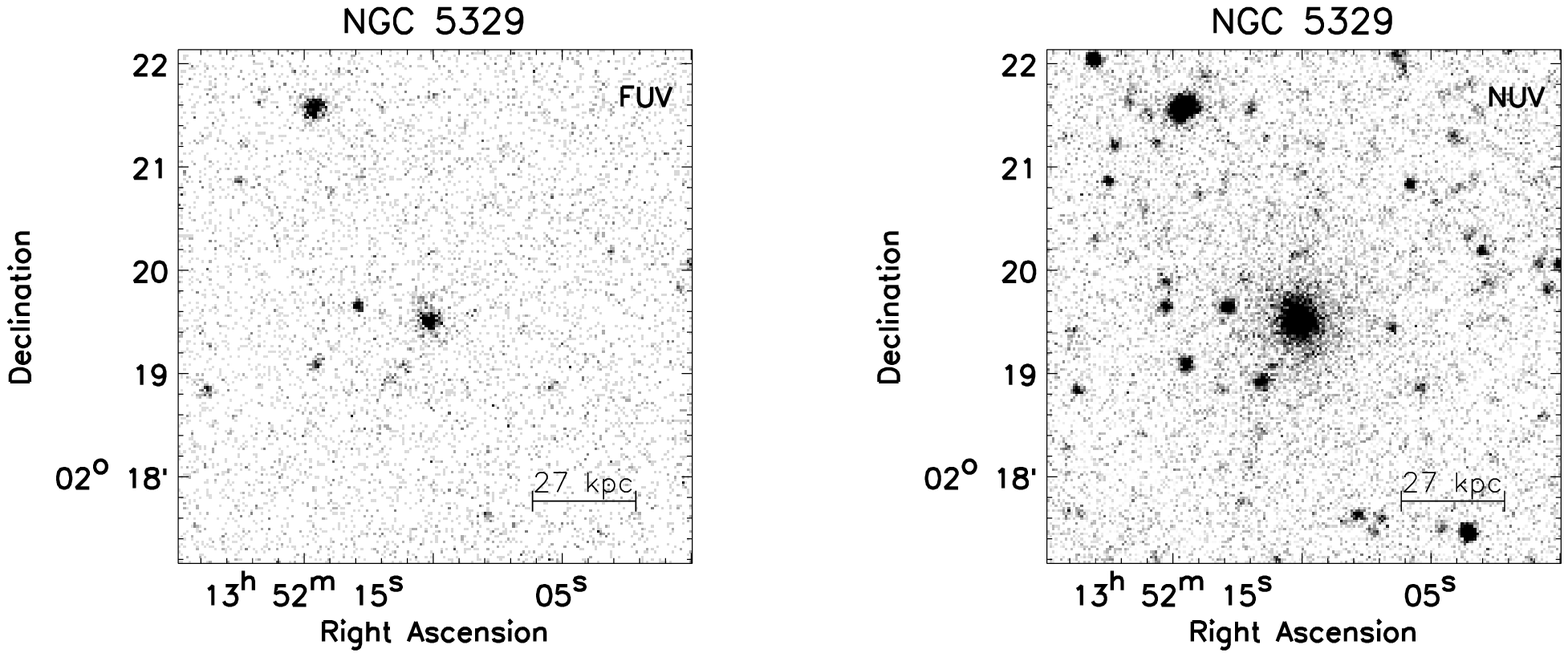}} 
\caption{Full resolution, 5\arcmin $\times$ 5\arcmin wide, FUV (left panels) 
and {\it GALEX} NUV  (right panels) background-subtracted images
(counts~px$^{-1}$~s$^{-1}$) of the three galaxies. The scale bar is
derived from adopted distances provided in Table~\ref{table1}.}
\label{fig1}
\end{figure*}

\begin{figure*}
\center
 {\includegraphics[width=7.5cm]{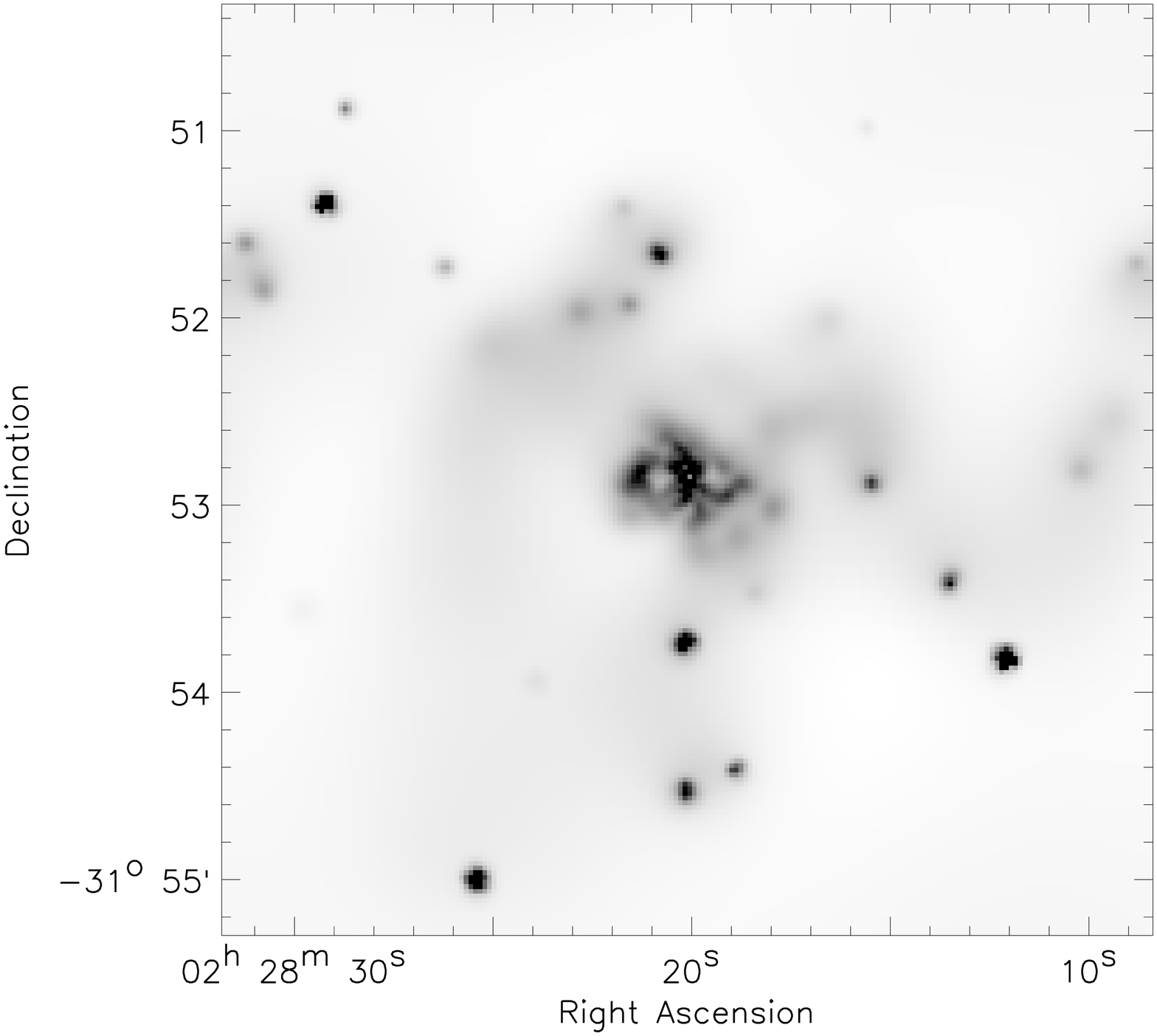}
\includegraphics[width=7.5cm]{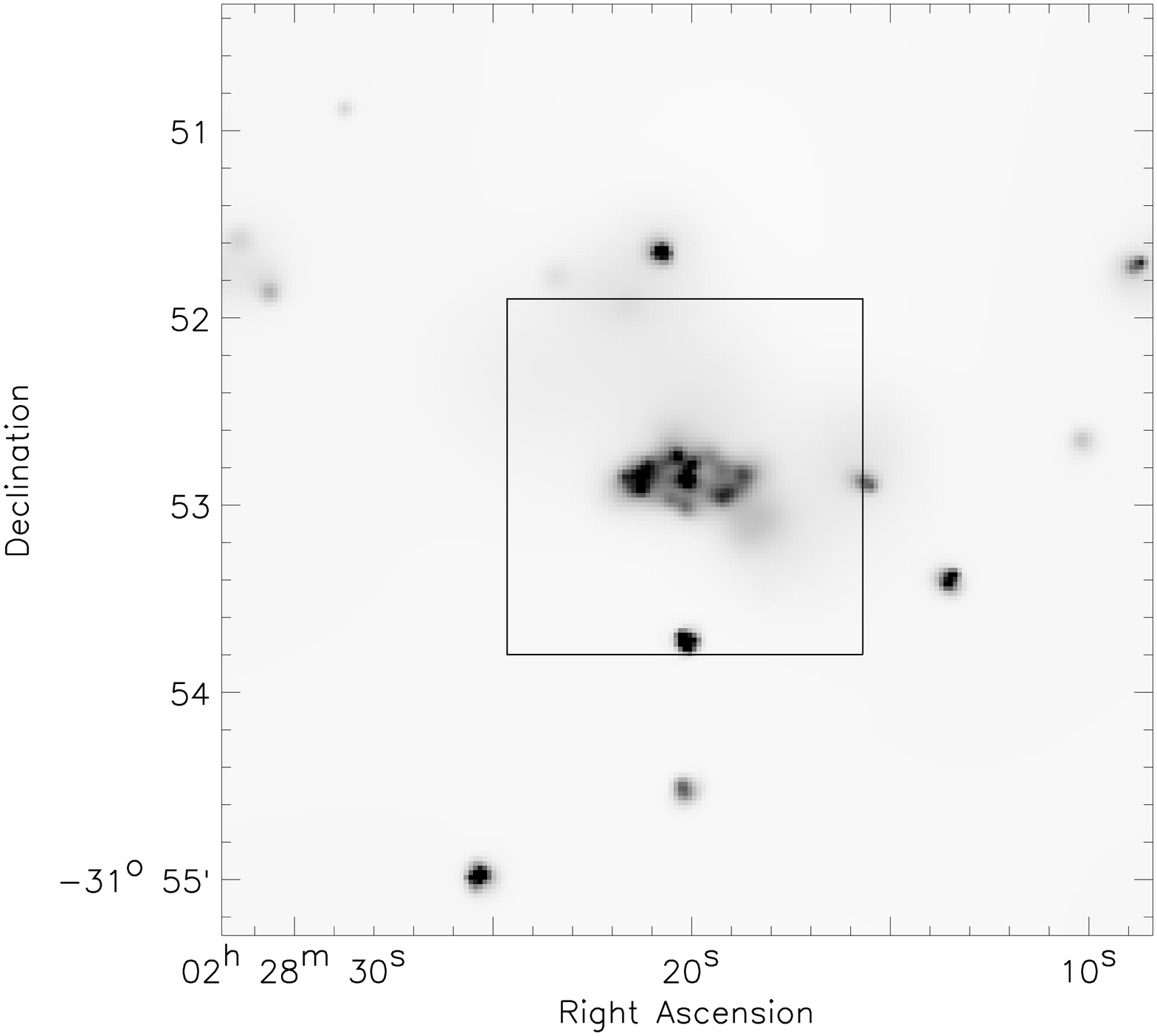}}
\includegraphics[width=5.7cm]{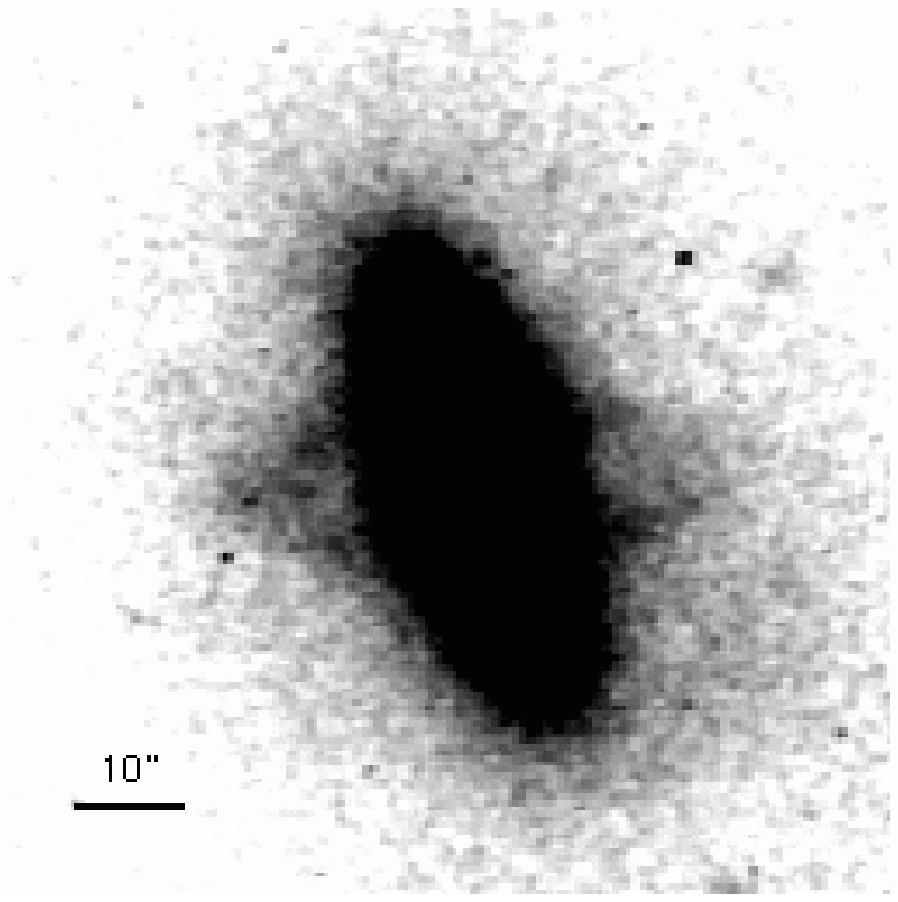}
\includegraphics[width=5.7cm]{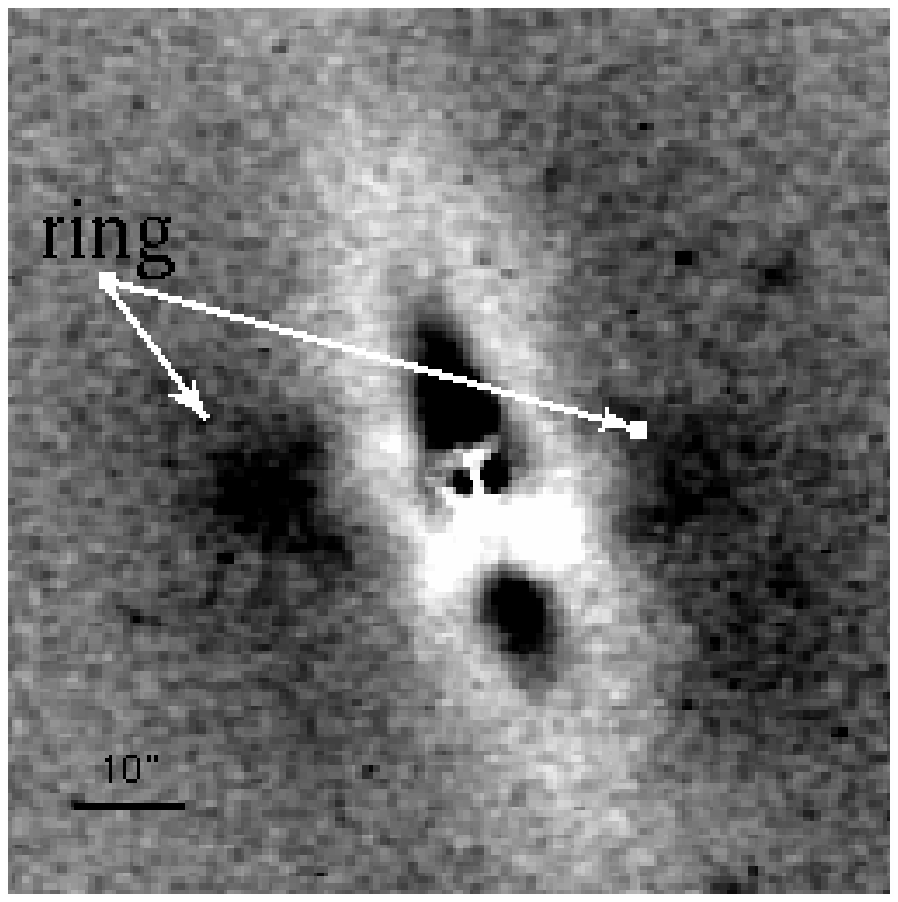}
\includegraphics[width=5.7cm]{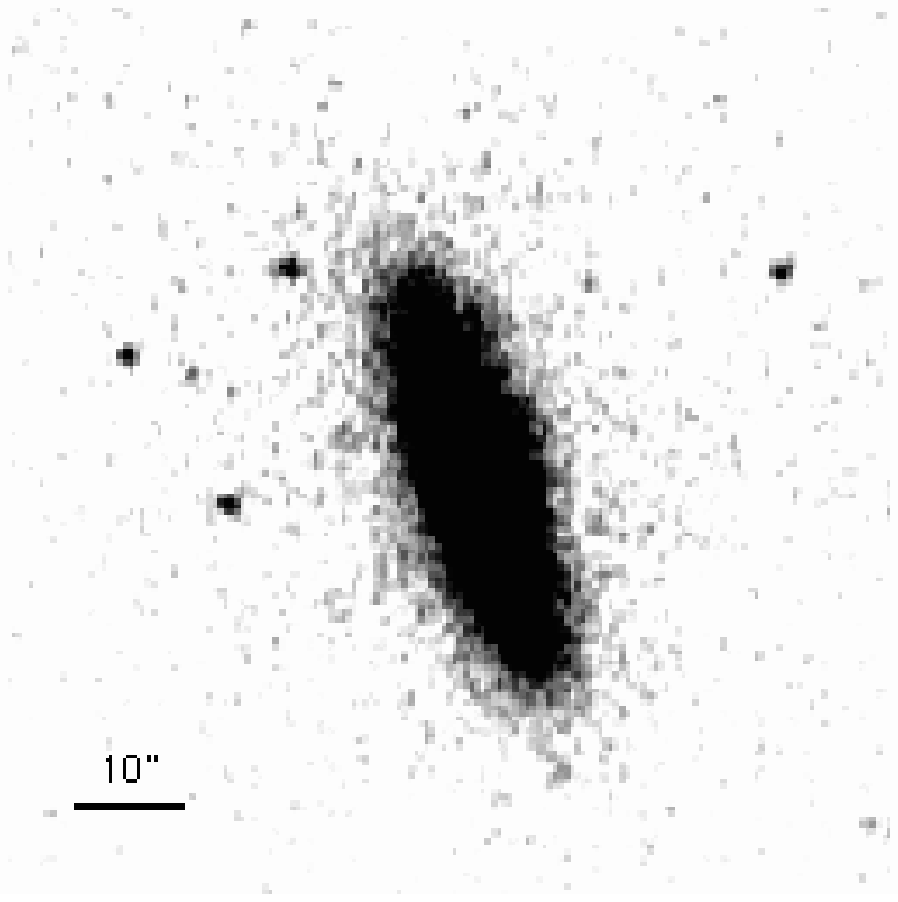}
\caption{ {\it GALEX} UV images in a 5\arcmin$\times$5\arcmin ~ field
around \MCGa. \textbf{ Top panels}: NUV and FUV images
obtained with ASMOOTH for $\tau_{min}$ = 5 and $\tau_{min}$ =3,
respectively. The insert  on the FUV image
indicates the field of view of the B images.
\textbf{Bottom panels}:  B-band image ({\it left}) and
B-band residuals with respect to  the 2D model ({\it middle}),
K-band image ({\it right}) \citep[see also][]{ Iodice02b}.  In the
K-band image, the polar ring is not visible, whereas  in  the B-band
and in the FUV and NUV bands the ring is the prominent feature. In the
B-band residuals, darker colours correspond to regions where the
galaxy is brighter than the model.}  
\label{fig2}
\end{figure*}

\begin{figure*}
\centering
\includegraphics[width=7.8cm]{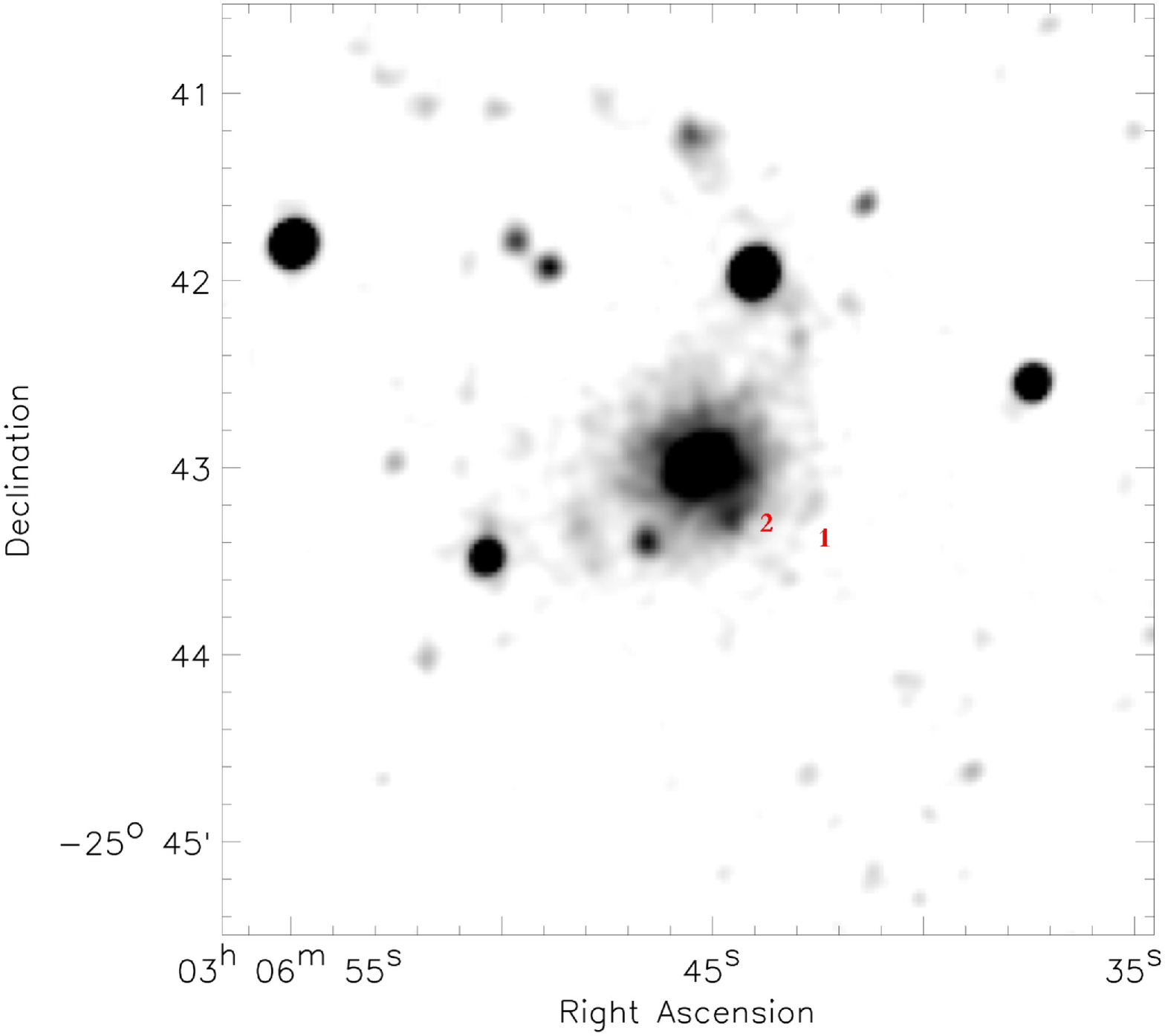}
\includegraphics[width=7.8cm]{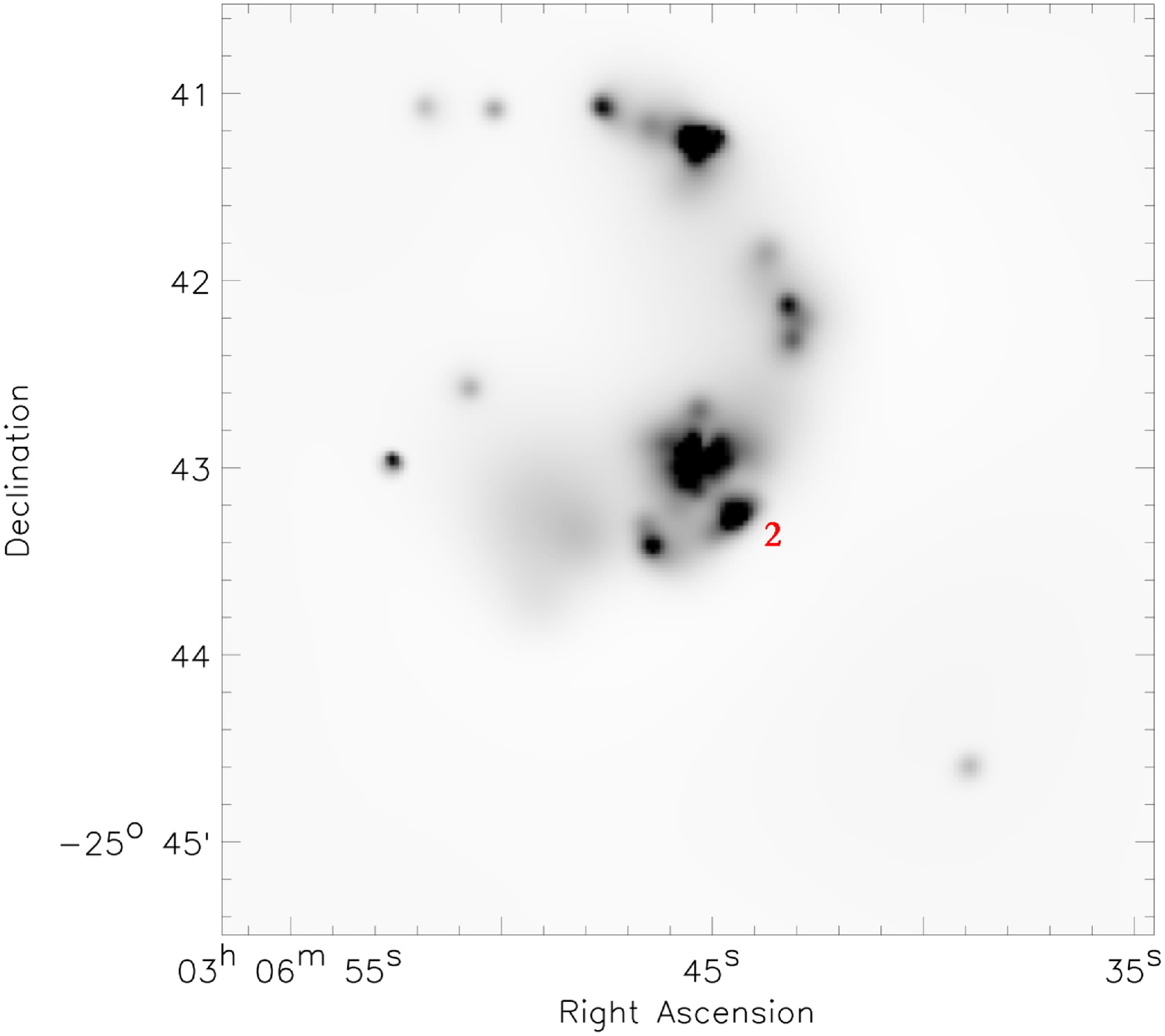}  \\
\includegraphics[width=5.7cm]{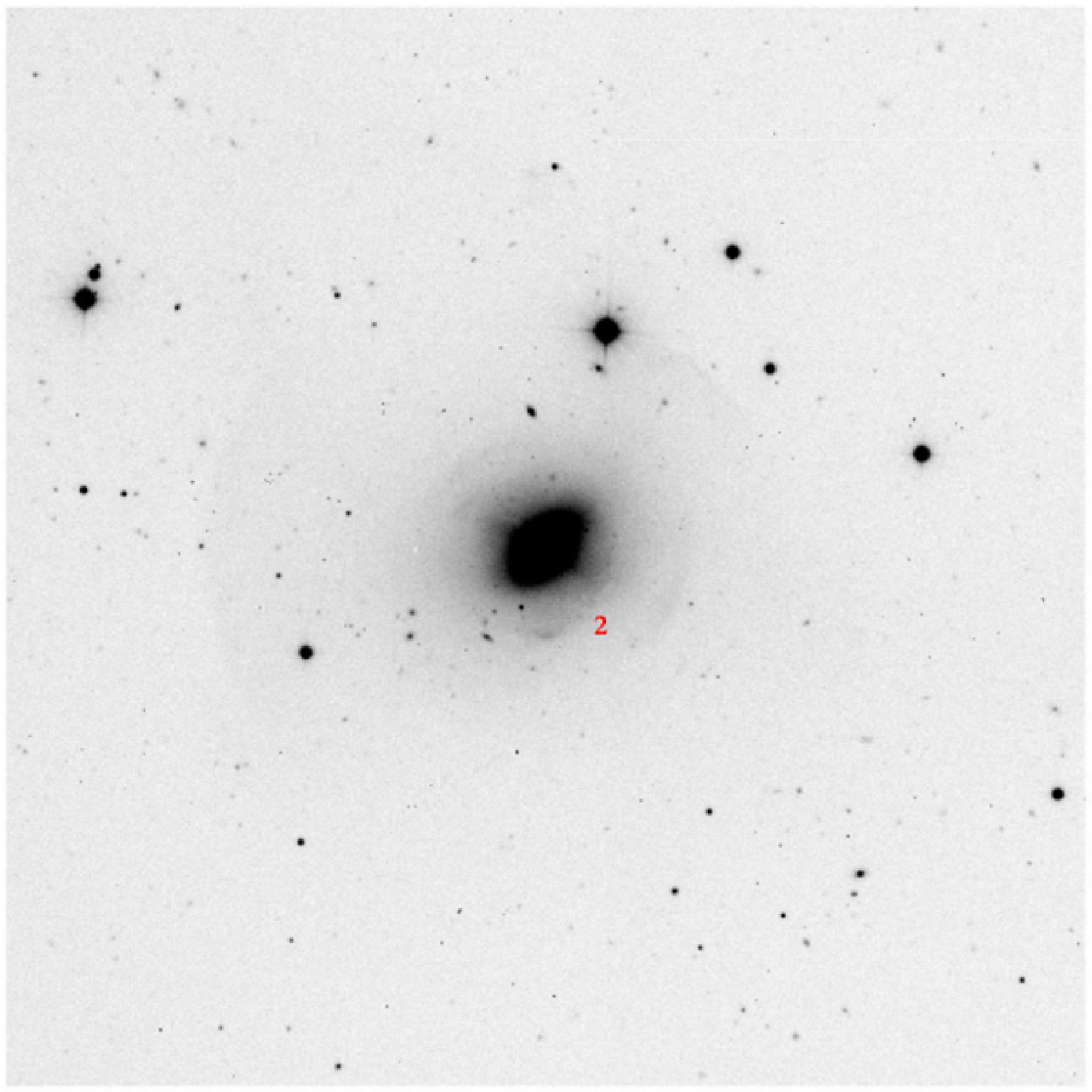}
\includegraphics[width=5.7cm]{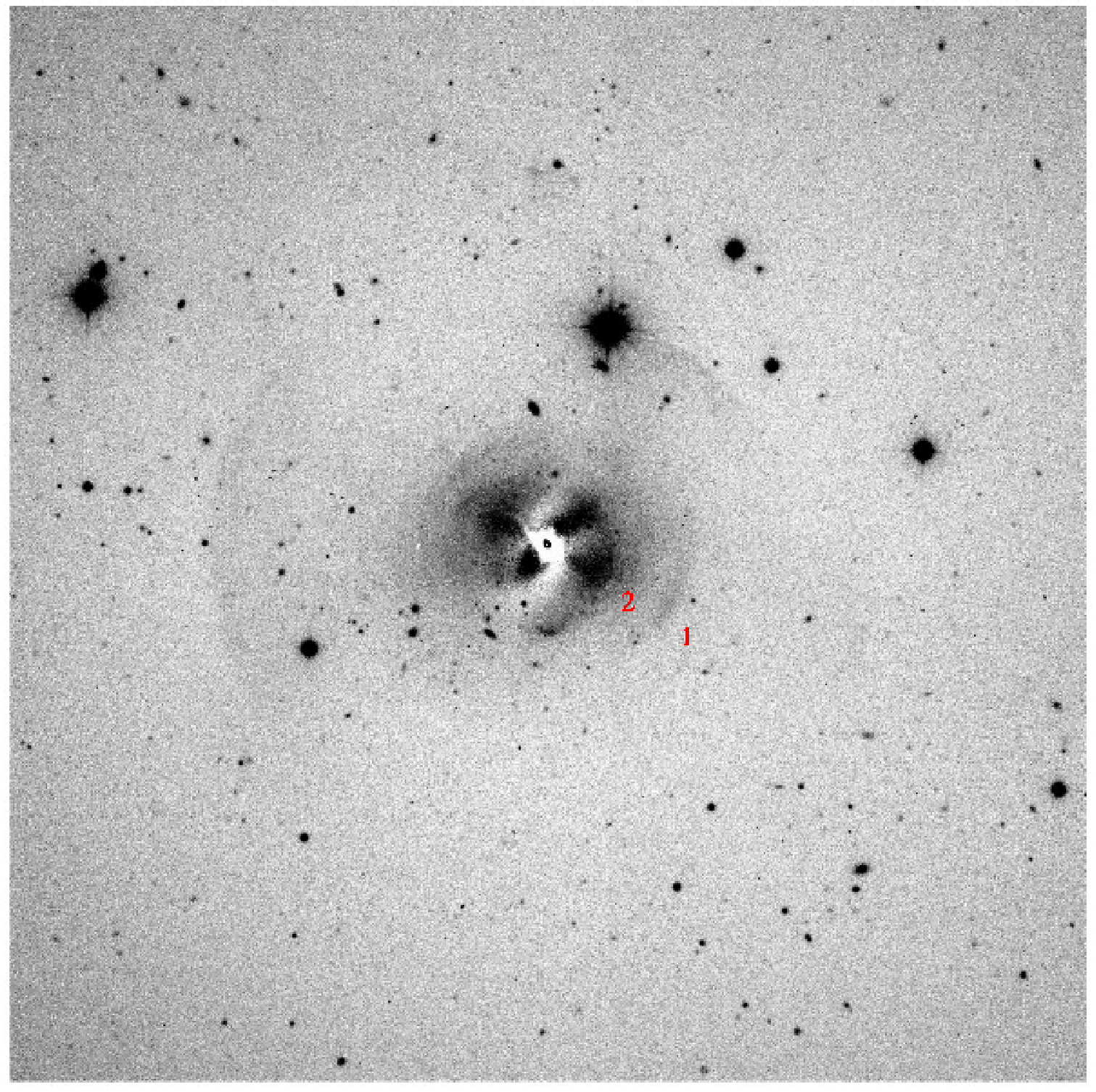}
\includegraphics[width=5.7cm]{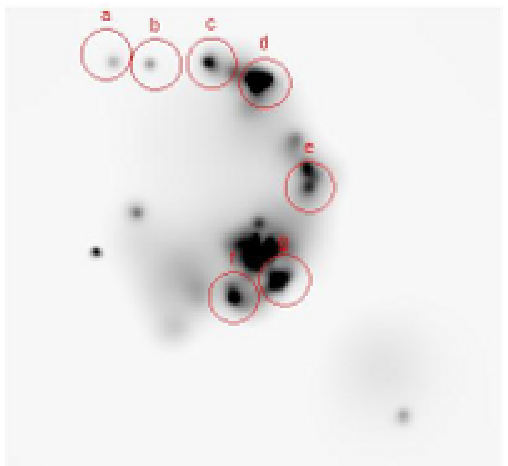}
\caption{ \textbf{ Top panels}: {\it GALEX} UV images in a 5\arcmin
$\times$5\arcmin~ field around NGC~1210. The NUV image (on the left)
was smoothed with an adaptive filter, whereas the FUV image (on
the right) with ASMOOTH ($\tau_{min}$ =3). Note how the system of
debris overlays the HI emission in the bottom panel of
Fig.~\ref{fig4}. \textbf{Bottom panels}: NTT/EMMI uncalibrated 
r-band image in a 5\arcmin $\times$5\arcmin (left) field and the same
image where the main body of the galaxy has been subtracted
revealing, together with shells, a complex central fine structure
(mid panel). Labeled ``1" and ``2" are the two shell edges visible in the
NUV and FUV images (see text).  Superimposed on the smoothed FUV image,
we indicate and label with letters the knots whose magnitudes are
provided in Table~\ref{table5} (right panel).} 
\label{fig3}
\end{figure*}

\section{The sample}\label{sample}

Table \ref{table1} presents the main physical
properties of our sample.

\textbf{\MCGa}, also known as ESO~415-G26, is a peculiar S0 galaxy
\citep{Whitmore87, Iodice02a}.  It is characterized by a polar ring
less extended than the central host galaxy in the optical band.
 Debris is present in the galaxy outskirts at a position angle that
is intermediate between the  major axis of the host galaxy and the
polar ring. \citet{Whitmore87} reported observations of a ``nice set
of ripples and an extended asymmetric envelope". In  attempting to date
the accretion event responsible for the formation of the ring, 
 they estimated an age of  1-3 Gyr.  The HI map of this
object was obtained with the VLA by \citet{vanGorkom87}, and
\citet{vanGorkom97}  noticed  that the neutral hydrogen lies
along the major axis of the polar ring, with some correlation
between HI and  outer shells. Based on the regularity of the HI
distribution, \citet{vanGorkom87} tried to estimate the age of the
event generating the ring. In brief, they argued  that the
timescale for evenly distributing HI in an orbital plane must be
proportional to the orbital time multiplied by the ratio of the
circular velocity to the typical random velocity. Since this last
factor is typically of the order of 10, whereas the ratio of the
Hubble time to the orbital time is more typically 100, they suggest
that the polar ring in \MCG is the result of a relatively
recent event. Optical data tend to support the timesscales derived
from the HI. In particular, the stellar debris in the vicinity of
\MCGa, may be indicative of an age younger than a few orbital times.

\textbf{NGC~1210} is classified as a peculiar and barred S0.  Its
near-infrared luminosity profile agrees with the Sersic law with
0.28$\pm$0.01$<1/n<$0.24$\pm$0.03,  i.e., is consistent with the
canonical $1/n$=0.25 corresponding to the de Vaucouleurs law
\citep{Brown03}. The galaxy kinematics and the line-strength indices
were measured by \citet{Longhetti98, Longhetti98a}. The high
velocity rotation (see Table~\ref{table1}) supports the idea that
the galaxy is a true S0. The H$\beta$ value provided by
\citet{Longhetti98a} is unfortunately not corrected for emission
infilling, although there is ionized gas ([OII]) in the galaxy
center, as shown in their Table~1.  The analysis by
\citet{Longhetti99} of the so-called ``blue line-strength" indices
(e.g., CaII[H+K] lines and $\Delta$4000) suggests that the galaxy has
a very young nucleus. NGC~1210 was observed in HI by
\citet{Schiminovich01a}. The HI emission forms a ring that includes
the galaxy center.

\textbf{NGC~5329} is a bright elliptical galaxy and the dominant
member of a poor cluster of galaxies, WBL~472, consisting of six
companion galaxies according to \citet{White99}.   In contrast, both
NGC~1210 \citep[see e.g.,][]{Malin83} and \MCG are isolated
galaxies. \citet{Brocca97} found that the nearest companion of \MCGa, is
PCG~9331, which is located at $\approx$17\arcmin\ (302 kpc) from the
galaxy with a velocity difference of $\approx$51 km~s$^{-1}$, well
beyond five times the diameter of the galaxy (6.1\arcmin), a
distance they consider as the limit for isolation.

To summarize, the galaxies in the present sample might plausibly represent
different aspects of a common phenomenon, i.e., an accretion
event. In the following paragraphs, we emphasize the similarities
and differences between these objects and those observed with {\it GALEX}
by  \citet{Rampazzo07}.

\MCG and NGC~1210 have a rich reservoir of HI in their main
body (see Table \ref{table1}), in contrast to that found for NGC~2865,
NGC~5018, and NGC~7135, the shell galaxies investigated by
\citet{Rampazzo07}, where HI is mostly detected outside the main
body of the  galaxies. HI is also found in other late-type  
members of the group, and in the tidal structures. \citet{Theureau07}
did not detect HI in the direction of the NGC~5329 group.

In the sample of \citet{Rampazzo07}, no galaxy displays Type 1
shells, i.e.,  systems of  aligned shells \citep[see e.g.,][\, for
definitions and statistics]{Prieur90}. NGC~2865 and NGC~5018 have
``all-round" shell systems that are known as Type 2; 7 shells are
detected in NGC~2865 and 10 in NGC~5018. While the type of  shell
system of NGC~5329 is unknown, \citet{Prieur90} classifies the shell
system of NGC~1210 as Type 2, and  finds  11 shells, containing
about 1.5-2\% of the total luminosity of the galaxy. Type 2 shells
are rather common as about 30\% of shell galaxies show this
structure \citep{Prieur90}. NGC~7135 \citep{Rampazzo07} has a Type 3
system of shells since they are neither aligned (Type 1) nor
``all-round" systems (Type 2). Following the description and images
in \citet{Whitmore87}, (see their Fig. 7 panels $c$ and $d$),  
\MCG\ should also be considered as having Type 3 shells. Indeed, shells
are seen in  the northeast
at 63\arcsec\ and 88\arcsec, and possibly in the West at 52\arcsec.
The major axis of the faintest material is misaligned at about 36
degrees with respect to the major axis of the S0 component.

In addition to the shells, \MCG presents a polar ring. In their Fig.
2 (mid panel), \citet{Rampazzo07} showed that the fine structure of
NGC~2865 does not solely consist of shells but also a tidal structure
(reminiscent of a ring) that crosses the entire galaxy in the east-west
direction, nearly perpendicularly to the major axis of the galaxy
and to the shell system. It is unclear whether this feature is connected
to the formation of the shells. In the case of the polar ring in
\MCGa, \citet{Whitmore87} suggested  that during the merger, the
stellar component of the accreted companion is dispersed into the
halo, whereas the gas clouds  experience collisions and quickly
settle onto a new disk or ring.

In this context, the analysis of the UV colours and  the
line-strength indices of NGC~2865 and NGC~7135 suggest that they both had
recent bursts of stellar activity in their centers, possibly as a
consequence of the interaction/accretion episode that triggered the
shell formation. \MCG and NGC~1210 probably experienced a major accretion
event. To cast light on these issues and in particular on the age of
the merger event,  with the aid of the  {\it GALEX} FUV and NUV
data, the optical line-strength indices,  the near-infrared
observations, and theoretical simulations of spectral energy
distributions, magnitudes, and colours, we try  to evaluate when in
the galaxy history the wet accretion (i.e., gas rich) events took
place. These estimates will be compared with timescales provided by
morphological and kinematical considerations given in the literature to
investigate  the possible connection between rejuvenation episodes
and the formation of the shell/rings.

\begin{table}
\caption{Journal of the {\it GALEX} observations}
\begin{tabular}{llllll}
\hline
\multicolumn{1}{l}{Ident.}&
\multicolumn{1}{l}{NUV}  &
\multicolumn{1}{l}{FUV}  &
\multicolumn{1}{l}{Observing} &
\multicolumn{1}{l}{P.I.} \\
\multicolumn{1}{l}{}&
\multicolumn{1}{l}{[sec]} &
\multicolumn{1}{l}{[sec]} &
\multicolumn{1}{l}{Date}&
\multicolumn{1}{l}{} \\
\hline
\MCG\,    & 1510   & 1531   & 2005-11-03  & D. Bettoni \\
NGC 1210  & 1558   & 1608   & 2005-11-04  & D. Bettoni  \\
NGC 5329  & 3869   & 2666   & 2004-06-09  & MIS \\
\hline
\end{tabular}
\label{table2}
\end{table}

\section{Observations } 
\label{observ}

The UV data were obtained with GALEX \citep[see][]{Martin05,
Morrissey05, Morrissey07},   a 50 cm diameter modified
Richey-Chr\`etien telescope with a very wide field of view (1.2
degrees) and a high image spatial resolution 
$\approx$4\farcs5 and 6\farcs0 FWHM in FUV (1350 -- 1750 \AA) and
NUV (1750 -- 2750 \AA) respectively, sampled with 1\farcs
5$\times$1\farcs 5 pixels.

NGC~1210 and \MCG were observed during dedicated runs assigned
to our team, whereas the data of NGC~5329 were taken from the
{\it GALEX} archive (Medium Imaging Survey: MIS). The observing
logs for each galaxy, including exposure times, are provided in
Table~\ref{table2}.  The full resolution images in the NUV and FUV
bands are shown in Fig.~\ref{fig1}.

 To the {\it GALEX} data, we add  the B- and K-band photometric data from
\citet{Iodice02a} for \MCG\, and  the analysis of the Sloan Digital
Sky Survey (SDSS) archival data in the  u [2980-4130 \AA], g
[3630-5830 \AA], r [5380-7230 \AA], i [6430-8630 \AA], and z [7730-11230
\AA] bands for NGC~5329 \citep[e.g.,][]{ Stoughton02}.

CCD images of NGC~1210 in the $Gunn~r$ band [6000-8500 \AA; central
wavelength 6733 \AA] were taken from the European Southern
Observatory, (ESO) Public Archive (prog: 68.B-0023). They were
obtained with EMMI at ESO-NTT telescope with the red filter R \#
773. The detector used for the observations consisted of two MIT/LL
CCDs. The average seeing had a FWHM$\approx$0\farcs84$\pm$0\farcs03
for the $Gunn~r$. 
Three exposures with this filter were available 
with an integration time of 200s. The raw frames were preprocessed
using the standard IRAF tasks for debiasing and flat-fielding. The
flat-field was performed using a normalized sky flat. Each CCD was
reduced independently: the images were dithered and shifted 
to obtain a complete image of  the galaxy and its surroundings.
After  applying debiasing and flat-fielding, the frames of each
CCD were combined to filter out spurious contamination by cosmic
rays.

\begin{table}
\caption{{Summary of the \it GALEX} photometric data }
\begin{tabular}{lccc}
\hline \multicolumn{1}{l}{}& \multicolumn{1}{c}{\MCG} &
\multicolumn{1}{l}{NGC~1210}  &
\multicolumn{1}{l}{NGC~5329} \\
\hline
r$_{e}^{NUV}$ [arcsec]         &   19           &   17            &  12              \\
r$_{e}^{FUV}$ [arcsec]         &   19           &   5             &   5              \\
m$^{tot}_{FUV}$(1530\AA)       & 19.78$\pm$0.07 & 19.46$\pm$0.08  & 20.20$\pm$0.04   \\
m$^{tot}_{NUV}$(2310\AA)       & 18.67$\pm$0.07 & 17.41$\pm$0.04  & 18.20$\pm$0.02   \\
(FUV-NUV)$^{tot}$              & 1.11$\pm$0.10  &  2.05$\pm$0.09  &  2.02$\pm$0.05   \\
\hline
\end{tabular}
\label{table3}
\end{table}

\section{Morphology and photometry}\label{results}

\subsection{Galaxy morphology}

\begin{figure}
{\includegraphics[width=8.5cm]{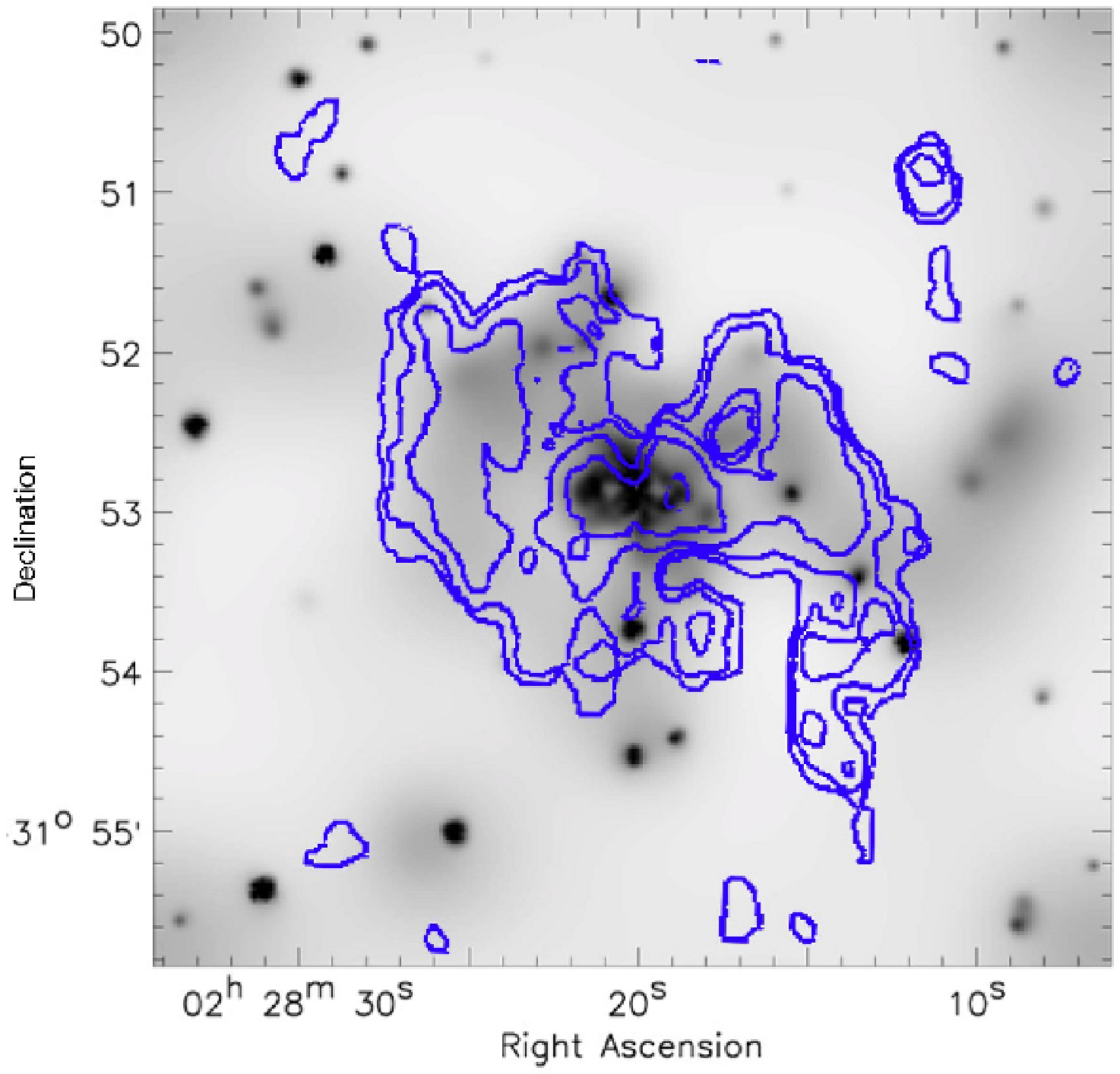}\\
\includegraphics[width=9.2cm]{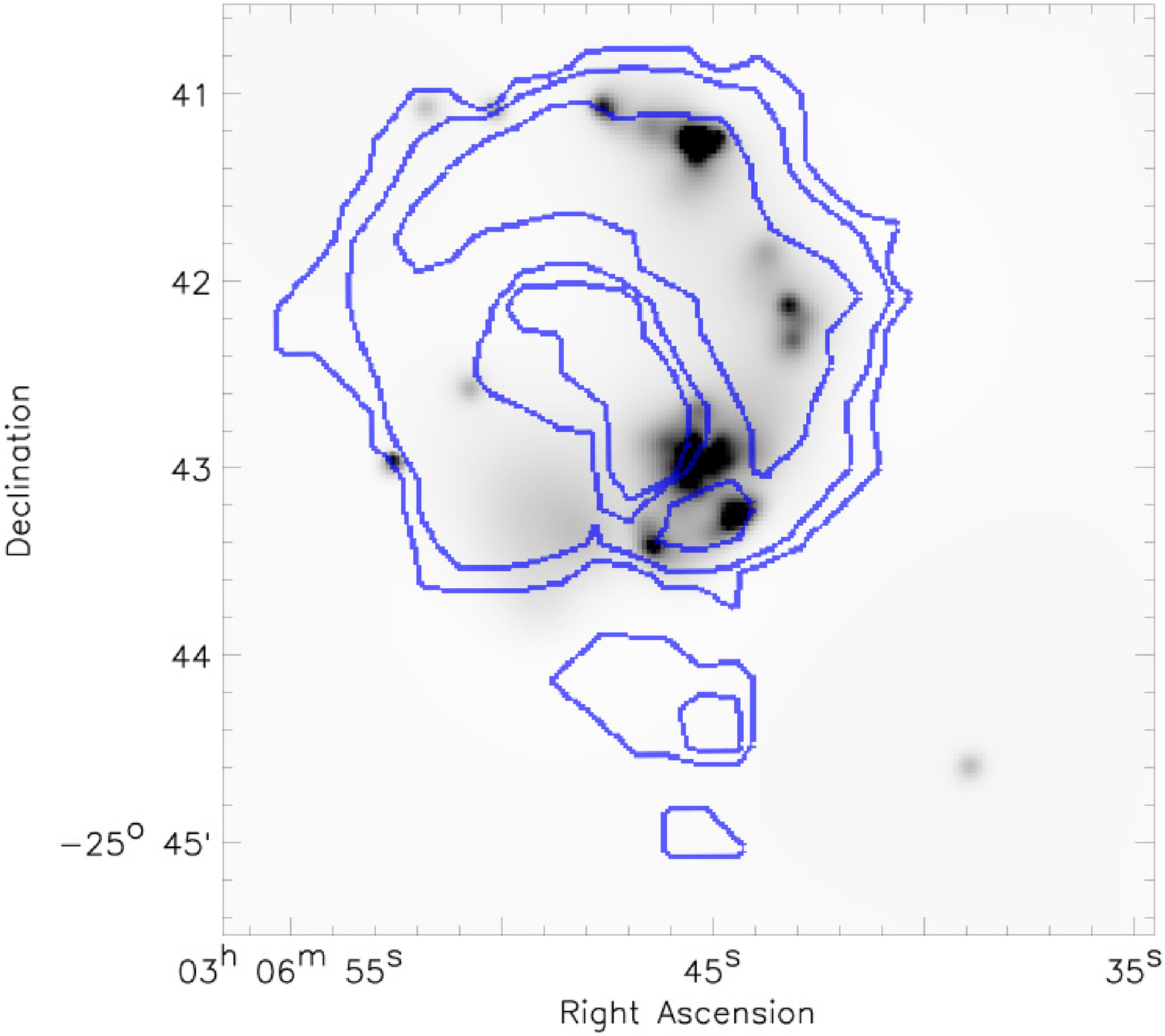}}\\
\caption{ \textbf{Top panel}: \MCGa. VLA B+C array
32\farcs1$\times$15\farcs2 resolution HI observations performed by
\citet{Schiminovich01} overlaid on the  FUV image.  The HI
emission overlays the polar ring and the system of debris visible in
the FUV image. \textbf{Bottom panel}: NGC 1210. VLA C+D array
52\arcsec$\times$36\arcsec resolution observations
\citep[see][]{Schiminovich01a}. Part of the HI ring  of NGC~1210 is
delineated by strong FUV emission. The polar ring in \MCGa,  and the
northwest part of the HI ring in NGC~1210 have blue
(FUV - NUV) colours suggesting very recent star formation.}
\label{fig4}
\end{figure}

\begin{figure}[!h]
\centering
\includegraphics[width=7.2cm]{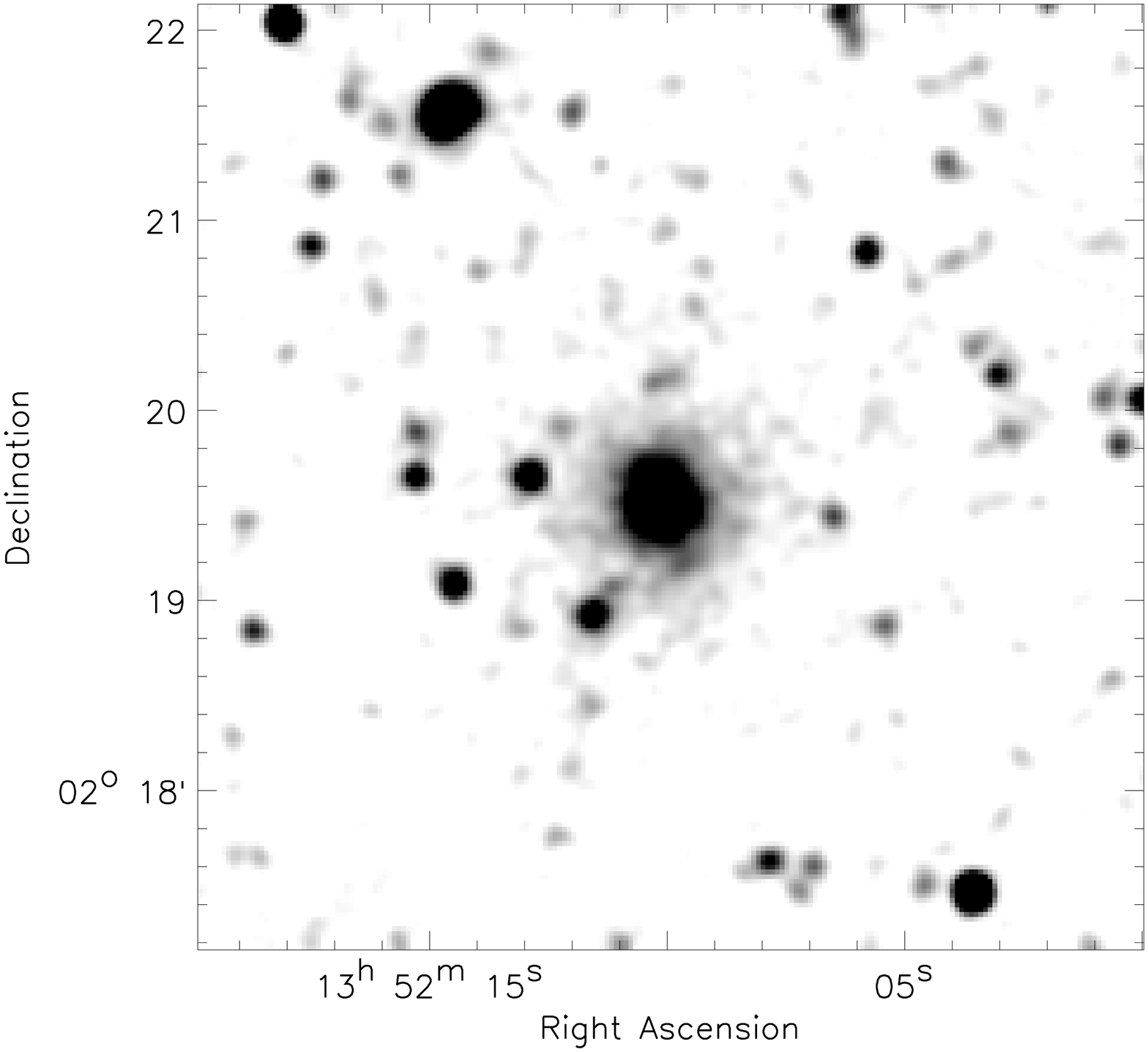}\\
\includegraphics[width=7.2cm]{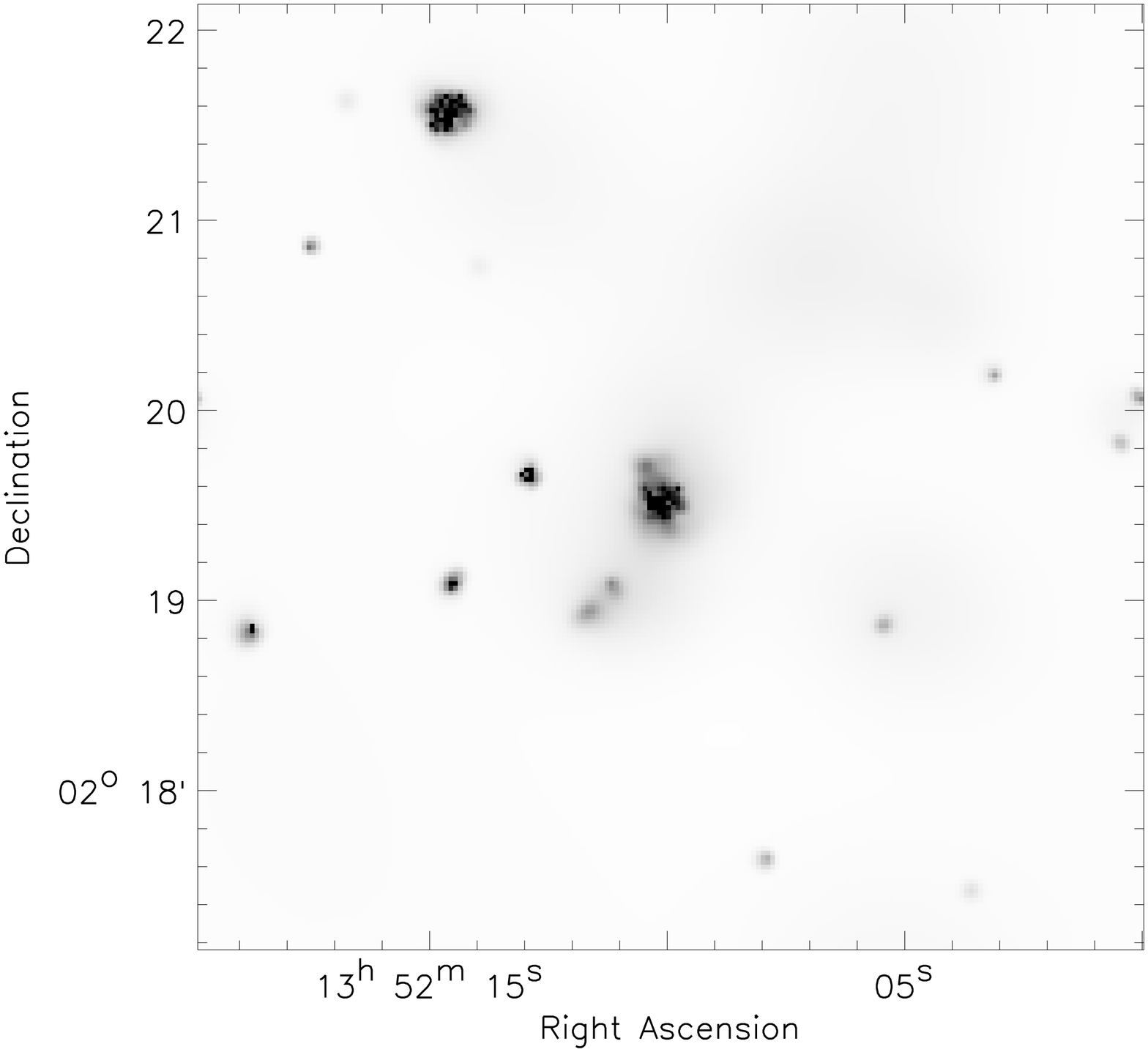}\\
\includegraphics[width=7.2cm]{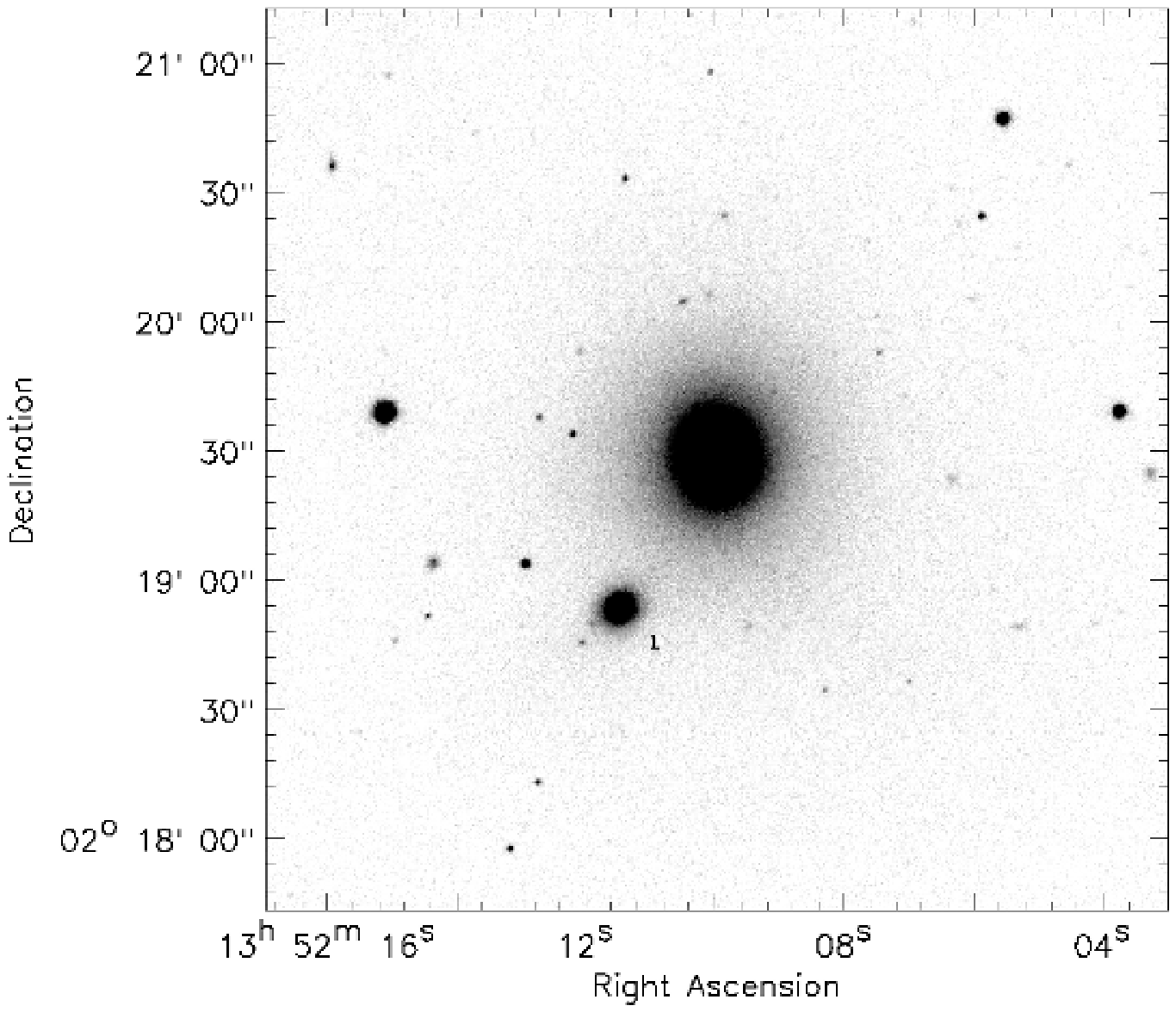}\\
 \caption{ {\it GALEX} UV images in a 5\arcmin $\times$ 5\arcmin~field
 around NGC~5329. \textbf{Top and mid panels}: NUV smoothed image and
 FUV image obtained with ASMOOTH for $\tau_{min}$ =3, respectively.
 \textbf{Bottom panel}: SDSS r image in 4\arcmin $\times$ 4\arcmin.
 The faint galaxy 2MASX J13521155+0218545 is physically associated with
 NGC~5329.} 
\label{fig5}
\end{figure}

To enhance the S/N in the galaxy outskirts and  bring out any
possible faint structures in the UV emission, we follow the procedure
outlined by \citet{Ebeling06}. The only  parameter required by {\tt
ASMOOTH} is the desired minimal S/N, $\tau_{min}$. For each
individual pixel, the algorithm increases the smoothing scale until
the S/N within the kernel reaches a specified  value. 
Thus, while the noise is suppressed very efficiently, the
signal (locally significant at the selected S/N level and carrying
information about the underlying physical cause) is preserved on
all scales. In particular, this  method allows us to
detect weak, extended structural  features falling in
noise-dominated regions (see Figs.~\ref{fig2}, \ref{fig3},
\ref{fig4}, and \ref{fig5}).

 \textbf{\MCGa}.~~  The morphology of this galaxy changes from
UV to NIR: the bulk of the NUV and FUV emission is concentrated in
the polar ring  (see Fig.~\ref{fig2} top panels) and the nucleus of
the galaxy. The main  body of the galaxy is visible in B- and
K-images (Fig.~\ref{fig2} bottom panels) but is hardly detectable
in  the NUV and FUV bands. In the B-band, both the central spheroid
and ring contribute to the light and are clearly detectable; only
the central spheroid contributes to the emission in the K-band;
finally, the outer ring  does not contribute  significantly  to the
K-band thus becoming invisible \citep[see also][]{Iodice02b}. After
applying  {\tt ASMOOTH} procedure to the NUV image, a large system
of debris appears. We draw attention to the features northeast and
northwest of the galaxy nucleus at a distance of about 1\arcmin\, which
lie at approximately the same distance and along the direction  of
the shell features detected by \citet{Whitmore87} (see their Fig. 7,
panels $c$ and $d$). An arc-like feature is also visible at
$\approx$30\arcsec\ from the center towards the southwest direction
in both the NUV and FUV images (in the latter, the feature is much
weaker). The same can also be seen in  panels $b$ of Fig.~7 in
\citet{Whitmore87}.
 We performed a 2D model of the B-light distribution in the
central spheroid using GALFIT and by masking  the outer regions of the
polar ring and foreground stars. The best-fit model implies that the central
object is a S0 galaxy with an exponential bulge. This result is
consistent with that derived by \citet{Iodice02b} using NIR images
(J, H, K). Subtracting the best-fit model  from the B-image, the
residuals are shown in Fig.~\ref{fig2} (middle panel): most of the
polar ring structure (on the southeast and northwest sides) is
 very clearly distinctive. On the southwest side, the polar ring passes in
front of the galaxy, thus the residuals show both the absorption by the
dust (lighter regions) and the emission by the stars in the ring
(darker regions). On the northeast side, where the ring passes
behind the galaxy, only the excess of light from the stars in the ring
is visible. Furthermore, the residual image detects 
an additional feature that encircles the entire galaxy and
corresponds to regions where the model is brighter than the galaxy:
this structure is also present in the NIR residuals (see Fig.1 in
Iodice et al. 2002b), where it has the same morphology but  is much
fainter, and in the NIR colour maps, where it appears bluer with
respect to the inner regions of the galaxy (see Fig.4 in Iodice et
al. 2002b). This ``cross-check" with the NIR images suggests that
this feature could be real and not the result of an artifact of the
fitting: it may indeed be an additional lens-like component.

\begin{figure}[!h]
\centering
\includegraphics[width=6.5cm]{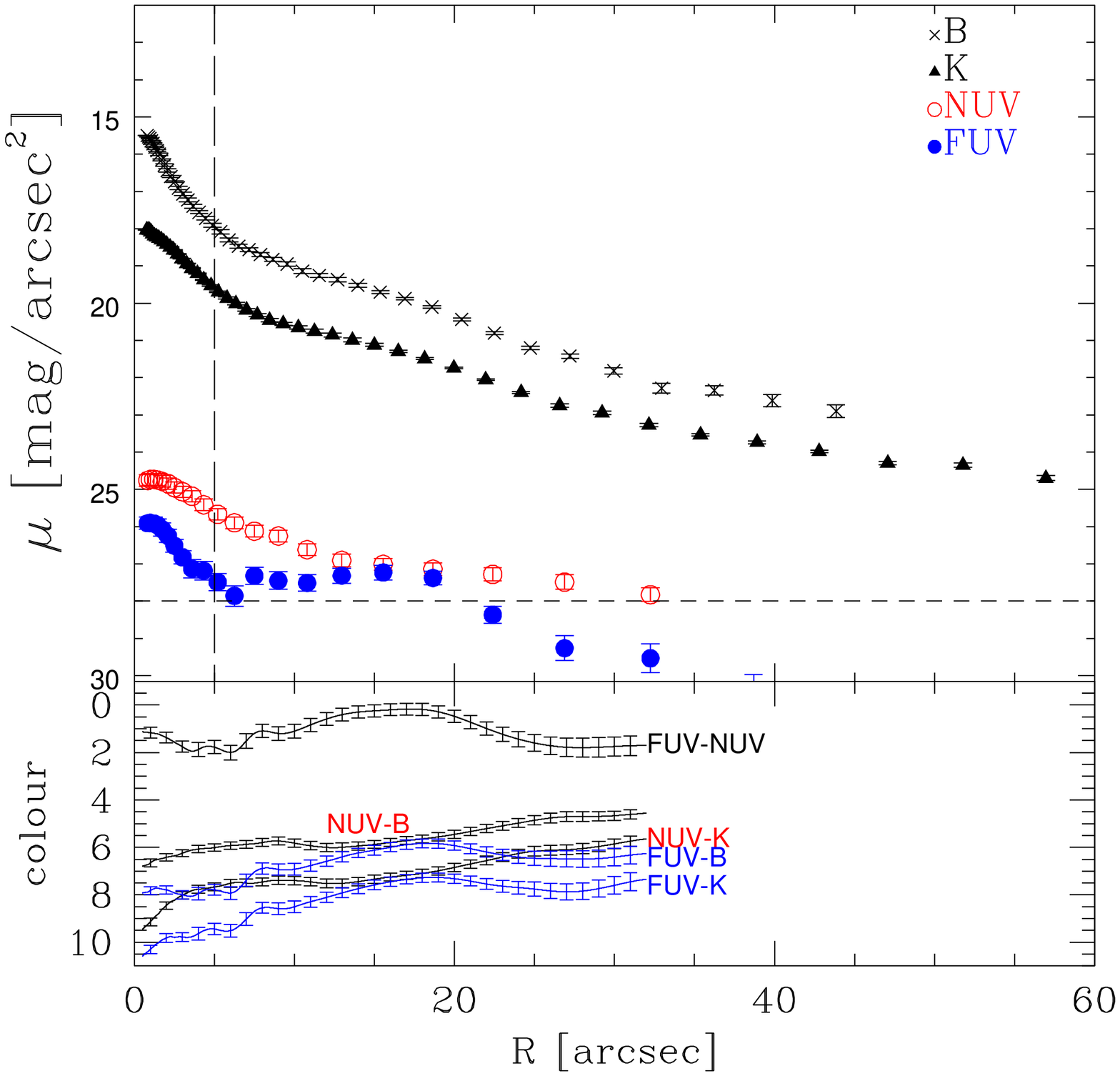} \\
\includegraphics[width=6.5cm]{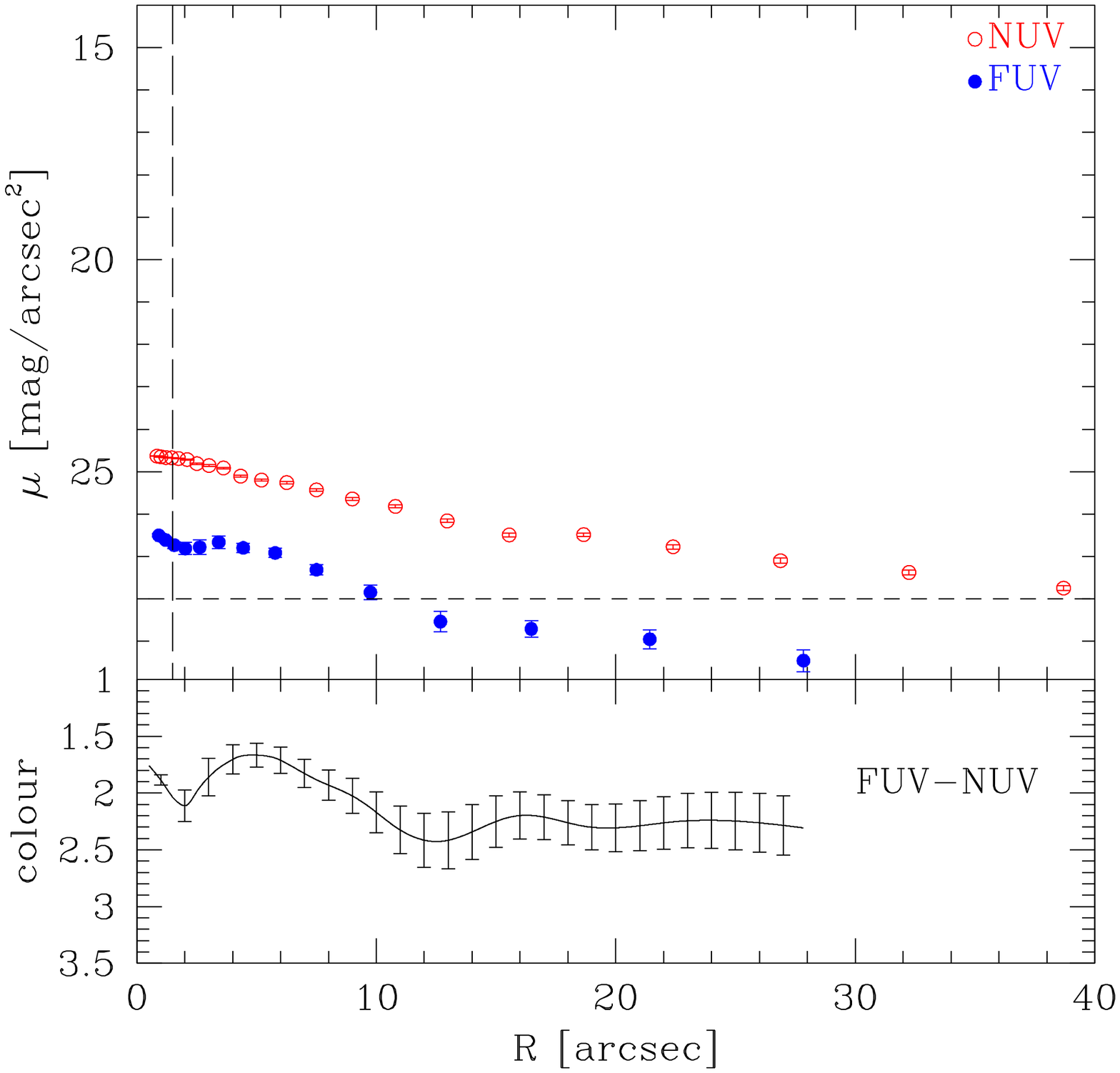}\\
\includegraphics[width=6.5cm]{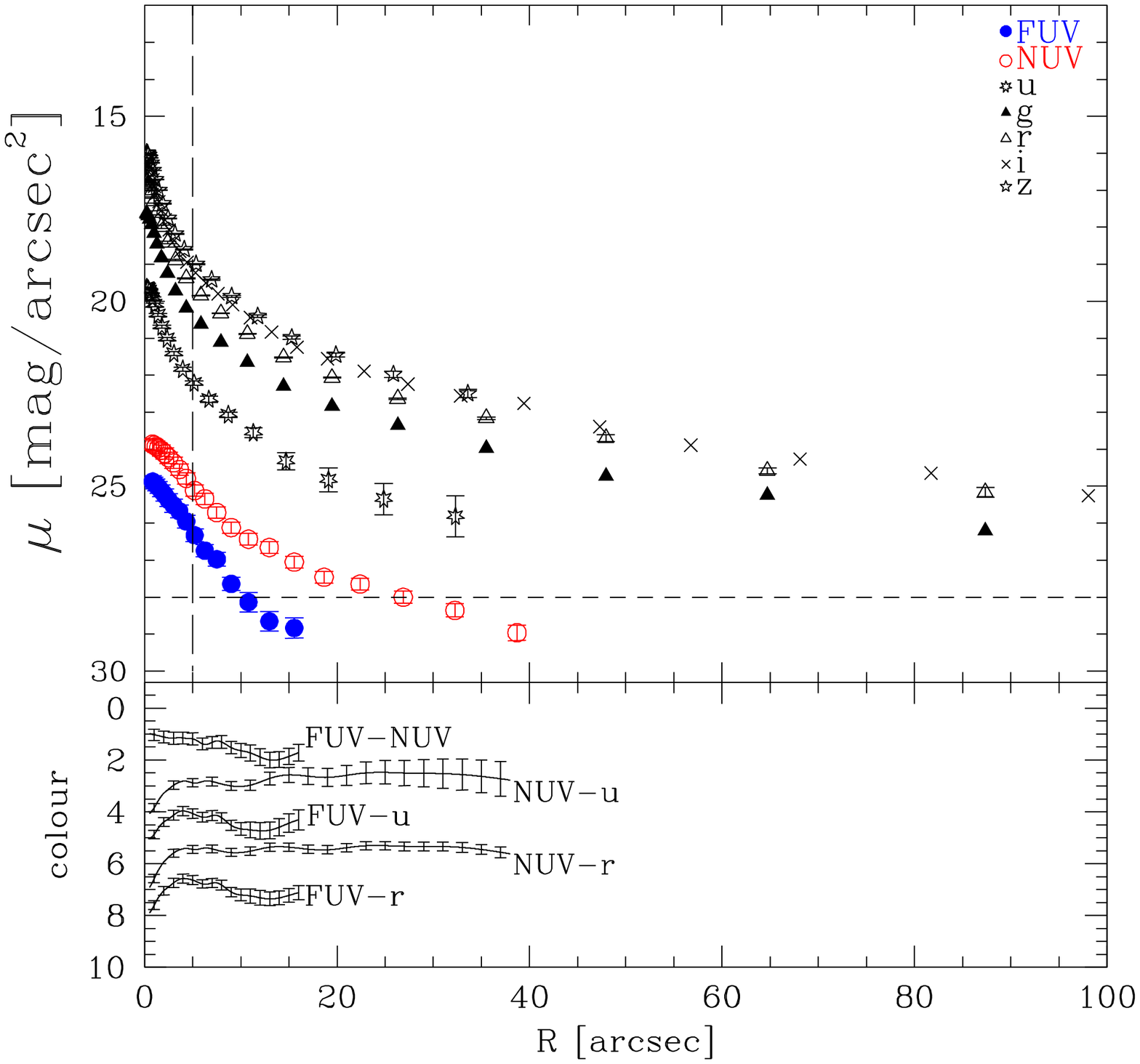}\\
\caption{ \textbf{Top panel}: {\it GALEX} UV, B and K band
\citep{Iodice02a} surface brightness of \MCGa.  \textbf{Mid
panel}: {\it GALEX} NUV and FUV surface brightness of NGC 1210.
\textbf{Bottom panel}: {\it GALEX} UV and SDSS u-g-r-i-z bands
surface brightness profiles of NGC~5329. The vertical dashed line at
5\arcsec\ indicates the approximate FWHM of the {\it GALEX} point
spread function. The horizontal line indicates the nominal UV
surface brightness limit.} 
\label{fig6}
\end{figure}

\begin{figure}[!h]
\centering
\includegraphics[width=8.2cm]{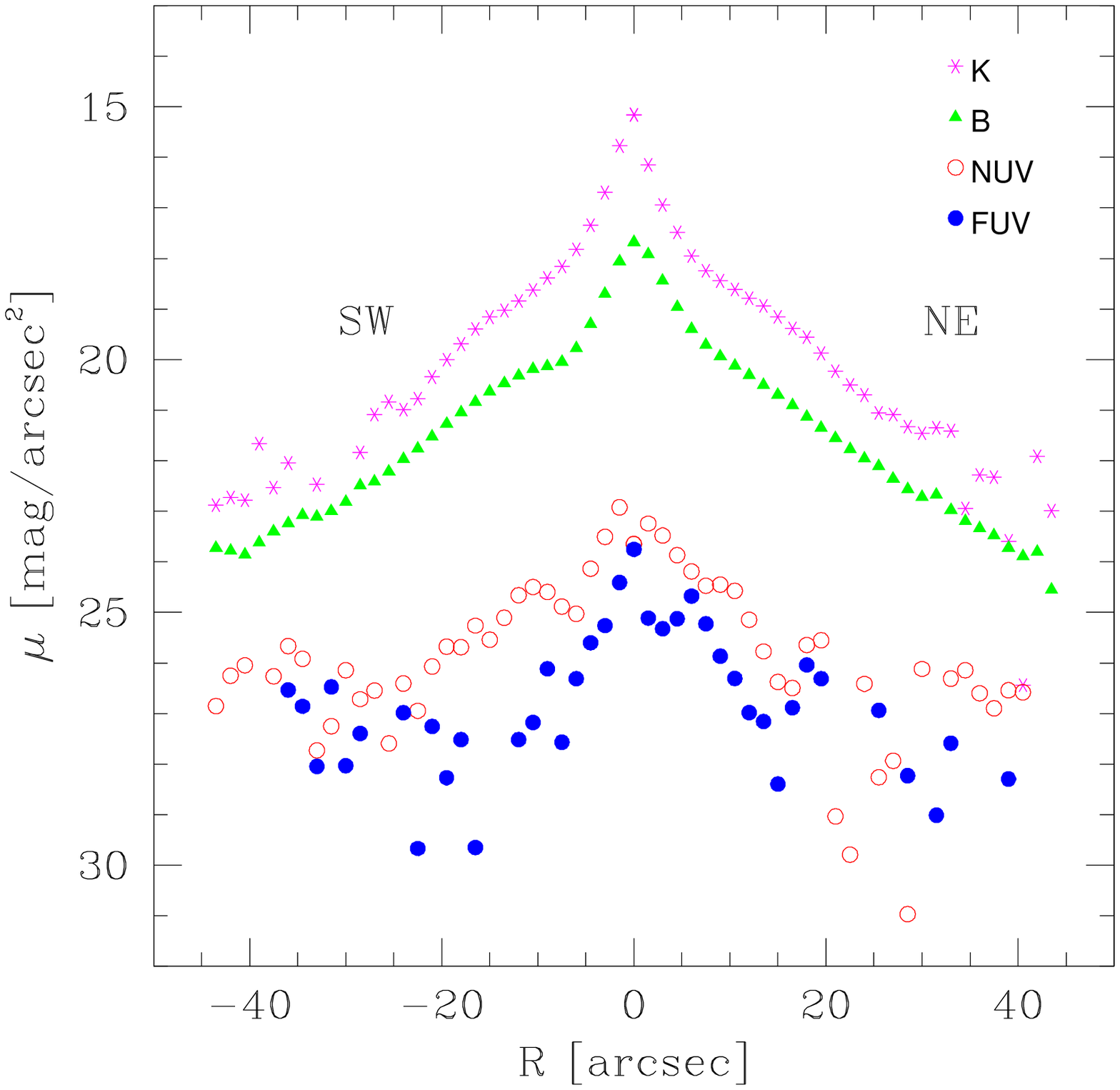}\\
\includegraphics[width=8.2cm]{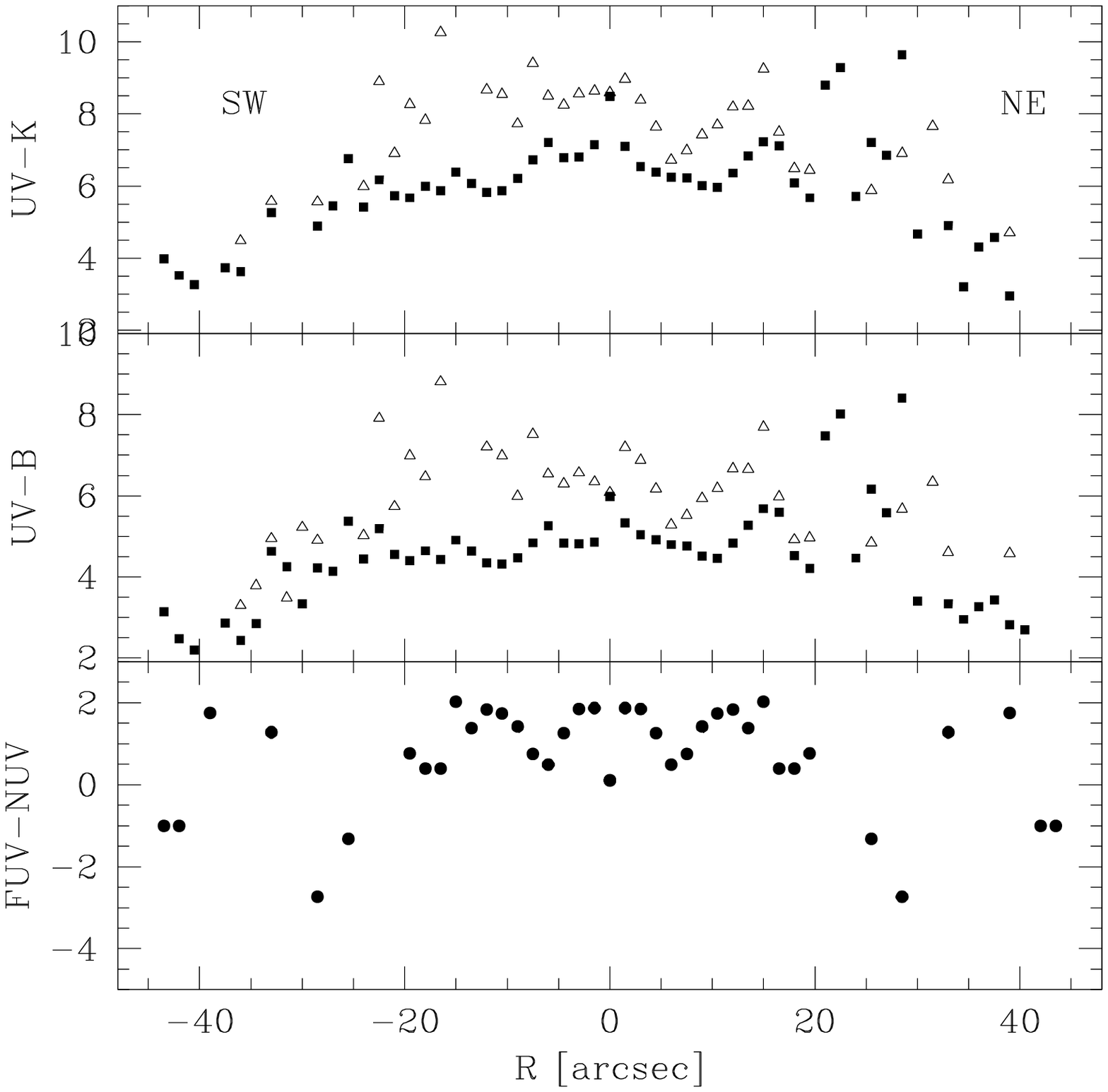}\\
\caption{ \textbf{Top panel}: UV, B and K light profiles along the
major axis of \MCG
 (P.A.=22$^{\circ}$).  The average uncertainty is 0.05 and 0.04 for B
 and K bands respectively.  \textbf{Bottom panel}: colours profiles along the
 \MCG\, major axis. The triangles indicate the FUV-B and FUV-K colours;
 the squares show the NUV-B and NUV-K colours. }
\label{fig7}
\end{figure}

\textbf{NGC~1210}.~~  As for the previous galaxy,  the NUV and FUV
images (Fig.~\ref{fig3} top panels) show a different structure
compared to that inferred from the  EMMI/NTT R-band image
(Fig.~\ref{fig3} left bottom panel). In the NUV and FUV passbands,
the galaxy appears to consist of a nucleus, from which a long
tidal tail departs toward  the north direction, and a diffuse cloud
of debris towards the south. Along this tail, several knots with a
luminosity similar to that of the nucleus are visible. In
the R band, the galaxy exhibits a system of shells, that are more
extended than the galaxy body. A faint, edge-on structure,
perpendicular to the major axis of the galaxy, is visible in the
inner part of the galaxy (see the left bottom panel of Fig.~\ref{fig4}).
Using GALFIT, we  performed a 2D model fit to the galaxy light distribution
 in the R band. Figure~\ref{fig3} (middle bottom panel)
shows the model-subtracted image, revealing a complex fine structure
composed of  shells, edge-on structures perpendicular
to the main galaxy body, and dust (as often found in shell
galaxies) \citep[see e.g.,][]{Sikkema07}.

The west edges of the two inner shells are labeled 1 and 2. Shell 1 is 
barely visible  in the NUV image. The FUV image has two main features: 
(i) shell 2 is more luminous than  shell 1 and (ii) there is a
long tail of stars and gaseous material, which seems to be part of
the same spiral-like structure extending towards the northeast. 
This suggests that shell 2 may have a different nature from
shell 1.
 
In Fig.~\ref{fig4}, we show the HI distribution of \MCG (top panel)
and NGC~1210 (bottom panel) measured by \citet{Schiminovich01}
and \citet{Schiminovich01a}, respectively. These are overlaid onto the
corresponding  FUV
images. For  \MCG, the HI distribution is elongated in the direction of
the polar ring. Furthermore, for  both \MCG and  NGC~1210 the HI
distribution coincides with the system of debris visible in  NUV
and FUV emission. In particular, in NGC~1210 the HI emission forms
a ring, which the tidal tail (detected by the {\it
GALEX} FUV observations), belongs to.

\textbf{NGC~5329}.~~ The NUV emission in NGC~5329 (Fig. \ref{fig5}, 
top panel) has an extensionsimilar to that of the optical image 
(Fig.  \ref{fig5}, bottom panel), whereas the FUV emission is 
concentrated in the central part of the galaxy (Fig.  \ref{fig5}, middle panel).
No peculiar structures, including shells, are detected in our
optical image.

To summarize, NGC~5329 has the
morphology of an elliptical galaxy in the NUV bands. In the same bands, 
NGC~1210 shows a long tail, which
is more clearly detected in the FUV band. In \MCG, the polar ring dominates both the
NUV and FUV emission,  whereas the
nucleus can hardly be seen in these bands.  A strong
correlation exists between both the FUV and NUV emission and the HI 
distribution in \MCG and NGC~1210.

\subsection{Light and colour distribution}

The surface photometry  of these galaxies is based  on  the background-subtracted
{\it GALEX} images and is calculated with  the {\tt IRAF-STSDAS-ELLIPSE} routine 
and  the {\tt GALFIT} package of \citet{Peng02}. The surface
photometry profiles are derived from {\tt  ELLIPSE}, which considers
the Fourier expansion of each successive isophote
\citep{Jedrzejewski87}. Figure~\ref{fig6} shows the surface brightness
and colour profiles of our galaxies. Furthermore, 
magnitudes, associated photometric errors, and
colours, are derived from the original unsmoothed data in the {\it
GALEX} AB magnitude system. The zeropoints of AB magnitudes are
taken from \citet{Morrissey07}.

The disturbed morphologies of \MCG and NGC~1210 render the
analysis of the surface brightness profiles more difficult than
usual, especially in the UV. The NUV and FUV luminosity profiles map
the nucleus and the ring of \MCG (Fig.~\ref{fig6}, top panel) and the
bulge of NGC~1210 (Fig.~\ref{fig6}, mid panel), respectively. In both
cases, a strong variation characterizes the profile of the  (FUV-NUV)
colour. The regular morphology of NGC~5329 has  a smooth
luminosity profile that follows the de Vaucouleurs $r^{1/4}$ law.

The (FUV - NUV) colour maps, not shown in this paper, indicate that
the nuclei of all the galaxies are red, with (FUV - NUV)$\simeq$1-1.5 or
even redder, probably because of  dust absorption. In contrast,
the knots in the polar ring of \MCG (Fig.~\ref{fig2}) and in the
northern tail of NGC~1210 (Fig.~\ref{fig3}) are  blue, (FUV -
NUV$\approx$0),  which is indicative of a young
component in the mix of the stellar populations. Table \ref{table3}
summarizes the  photometric data in the far UV and the structural
properties of the three galaxies.

The B and K luminosity profiles of \MCG are much more
extended than the UV light profiles, since this light is produced by  
the entire body of the galaxy.  Figure~\ref{fig7} compares the
B, K, and far-UV light profiles along the major axis of the
main body of the galaxy, i.e., P.A.=22$^{\circ}$.
Colour gradients are visible not only in direction of the ring
(P.A.=94$^{\circ}$) but in general from the centre to the periphery of
the galaxy, since the nucleus is always redder than the outskirts.

The UV and optical colour profiles of NGC~5329 (Fig.  \ref{fig6} bottom panel) 
are quite regular. Except for FUV-NUV, the nucleus is bluer in the other
colour profiles than the galaxy outskirts. Given
(NUV-r)=5.6$\pm$0.02 and $M_r=-22.05$, NGC 5329 is located on the
red sequence of the (NUV-r) versus $M_r$ colour-magnitude diagram of
\citet{Salim07} (see their Fig.~1).

\begin{figure}
\centering {\includegraphics[width=9.0cm]{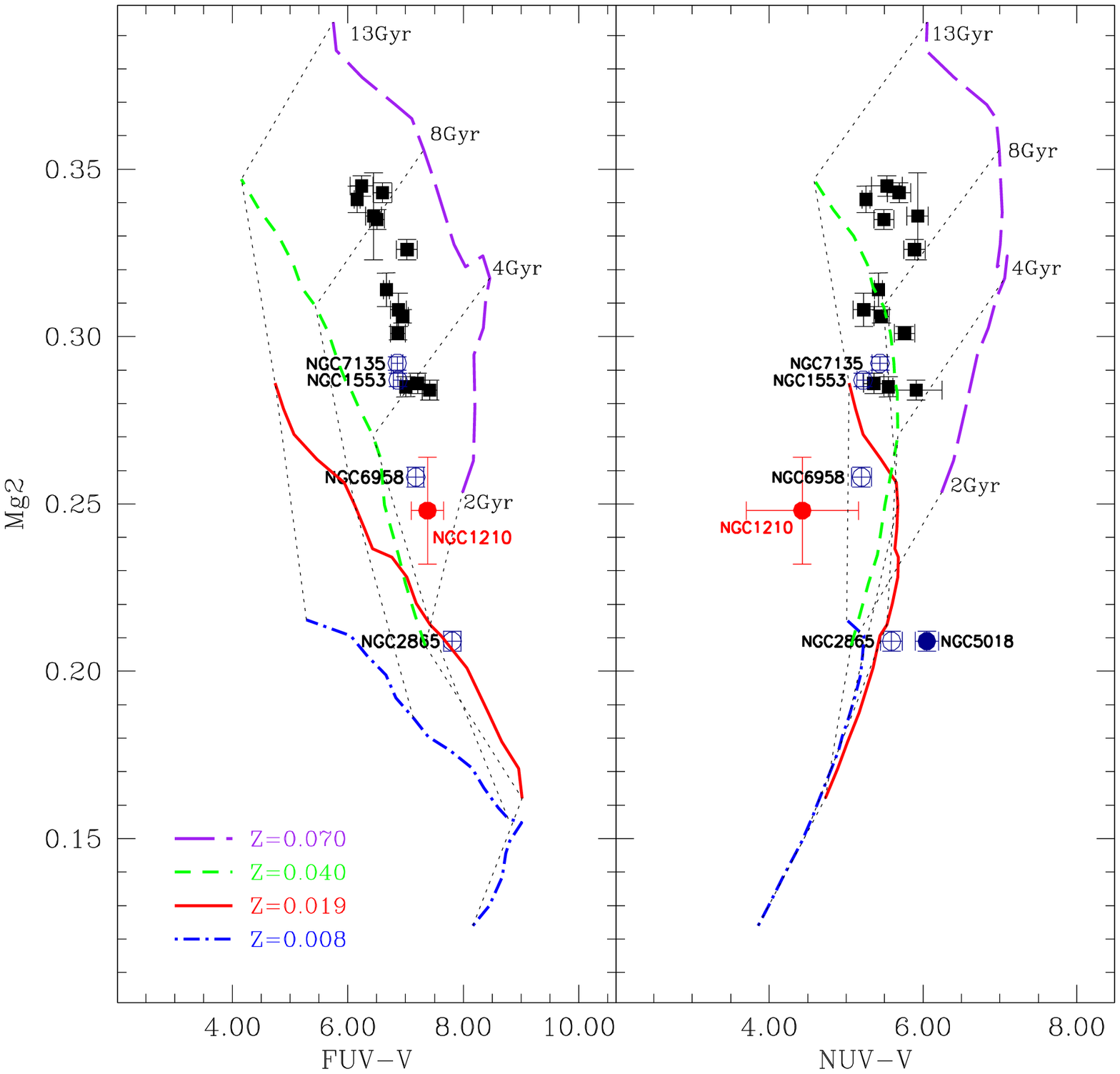}}
\caption{Comparison of the theoretical relation Mg2 versus {\it GALEX}
FUV (left) and NUV (right) - optical colours with observations.
Dust-free SSPs of different age and metallicity are used. The loci
of constant age and metallicity are shown as indicated. The  data
are from \citet{Rampazzo07}.}
\label{fig8a}
\end{figure}

\section{Theory versus observations: a possible interpretation}\label{theo_vs_obs}

\subsection{How recent are the accretion events?}\label{SSPanalysis}

The UV colours are  particularly suited to detecting
very young stellar populations inside interacting galaxies
\citep[see e.g.,][]{Hibbard05}. However, some issues, need to be
discussed when dealing with UV colours.

Firstly, the possibility of high-metallicity, old stars (age $\geq
10$ Gyr and metallicity $Z\simeq 3 Z_\odot$) being present in
early-type galaxies, even in small amounts, and could affect the
(FUV-NUV) colours. Indeed, the evolved stars in metal-rich
populations (extremely hot-HB and AGB-manqu\'e stars) are  
widely accepted sources of the UV excess in quiescent
early-type galaxies \citep{Bressan94}. Therefore,  all models
designed to reproduce the (FUV-NUV) colours should consider
that these high-metallicity stars are present, in
particular when folding together young and old stellar populations.
If some recent star formation activity adds young stars to a
dominant  older  stellar population making the UV colour 
blue, one should,  before drawing any conclusion about the young stellar
component, correct the UV colours for the contribution of
very old stars, which will not be negligible.

Secondly,   our galaxy sample could exhibit  strong morphological
peculiarities such as   polar rings and shells, customarily
attributed   to recent accretion/merging episodes that could have
induced some star formation \citep[see e.g.,][]{Longhetti00,
Rampazzo07}. It is then natural to believe that their stellar
populations should span a range of ages, i.e., young stars
could be present and thus  contribute, in different proportions, to
both the NUV and FUV fluxes.

Additionally    in star-forming regions, the UV-optical
stellar fluxes are strongly obscured by dust. The radiation in the
UV-optical range is absorbed, and the effect is stronger at shorter
wavelengths \citep{Draine03}, and re-emitted at
 longer wavelengths in the MIR/FIR. Therefore, dust  can strongly
affect the observed UV/optical magnitudes and colours and alter the
shape of the SED. Modelling galaxies with star-forming regions 
requires an accurate treatment of  dust, because of the combined
extinction/emission effects.

Finally, the comparison of data with theory is made assuming that
the complex stellar mix of a real galaxy can be reduced to a SSP of
suitable metallicity and age. However, this approximation has
different implications for the two parameters. While the metallicity
distribution can be ``reasonably" approximated to the mean value, the
same does not hold for the age, when this value is derived from integrated
properties \citep[see e.g.,][]{Serra07b}. In the discussion below,
one has to keep in mind that {\it the age that we are measuring from
colours (and/or indices) is always biased by the last episode of star
formation}. In other words, it is a mean luminosity-weighted age, in
which the most recent star-forming episode dominates at least during
the first $2-3$ Gyr from its occurrence. This is simply because of the
rather well established law of luminosity fading of stellar
populations, which ultimately mirrors the lifetime and evolutionary
rate of a star as a function of its mass.  So the ages derived
adopting theoretical models represent lower limits to the age of the
star-forming event (by accretion or whatever).

With the above points in mind, we apply the population synthesis
technique to interpret   our observational data and infer
the secular evolution of our galaxies.

\begin{figure*} 
\centering
 {\includegraphics[width=14.0cm,height=14.0cm]{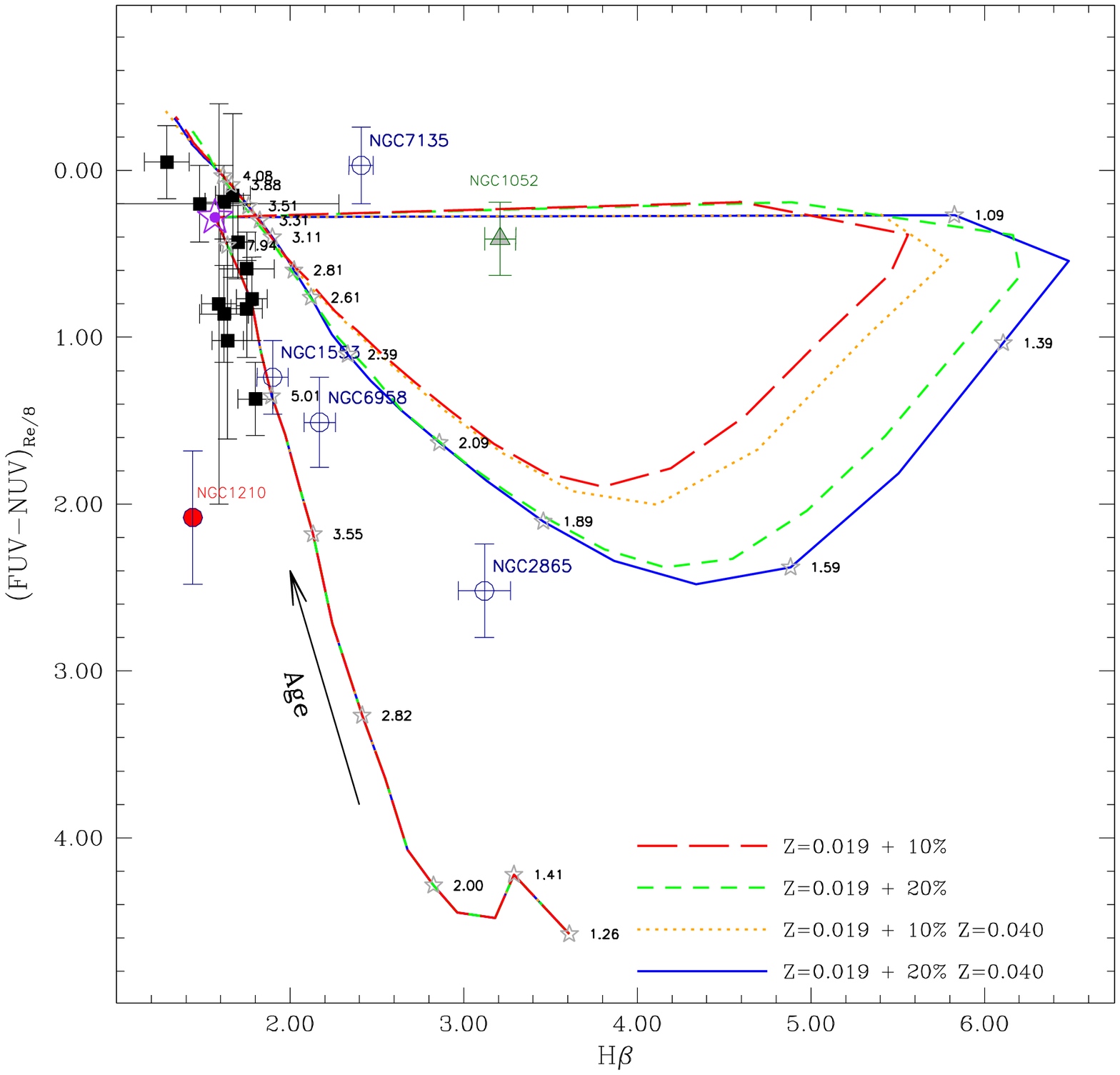}}
 \caption{ 
 {\it GALEX} (FUV - NUV) colours, within an aperture of r$_{e}$/8
radius, versus H$\beta$ of the sample of \citet{Rampazzo07} and of NGC 1210
(red bullet). We also show the evolution of a SSP as a function of time 
(solid red line along which the arrow
labelled age is drawn). The age increases as indicated by the arrow
and a few values of the age in Gyr are marked.
The position of a 10 Gyr old galaxy of solar chemical composition is indicated by
the large star located in the top left part of the plot. A burst
of star formation, superimposed on such old stellar component,
produces the "closed paths" traced in the plot by dotted, longdashed, 
short-dashed, and solid lines according to the percentage
of the mass involved in the burst and of the different metallicity
in the newly born stars.
The path starts from the
empty stars at the top left, performs an extended loop, and goes back
to a new position very close to the initial one. Several values of
the age since the beginning of the burst are indicated. The true age
of the composite SSP is $10 + T_{burst}$ Gyr.  Four combinations of the
metallicity $Z$ and  the percentage of star mass  in the burst are considered 
as shown by the insert. 
Indeed a burst of activity can deeply change the UV colour and the $H\beta$ index of
an underlying old population. Notice that the H$\beta$ value was not
corrected for emission-line infilling: the correction for the
presence of an emission component within the absorption line would
move the H$\beta$ line-strength indices toward higher values (right
side of the plot) toward the location of NGC~2865.  The Mg2 and
H$\beta$ line-strength indices values of NGC~1210 are taken from
\citet{Longhetti00}. }
\label{fig8b}
\end{figure*}

In the present study,  only NGC~1210 has measured values of Mg2
(0.248$\pm$0.016) and H$\beta$ (1.44) line-strength indices
\citep{Longhetti00}. The broad-band colours estimated for NGC~1210
are (FUV-NUV)=2.09$\pm$0.40, (FUV-V)=7.38$\pm$0.28, and
(NUV-V)=4.43$\pm$0.73.

In the case of \MCG and NGC~5329, we must adopt a different
strategy based on the modelling of UV {\it GALEX}, optical, and NIR
colours measured in suitable apertures. For \MCG,  we computed
magnitudes within three apertures, of $r_e$/4 radius, located on the
galaxy nucleus and north and south of the galaxy nucleus along the
major axis (P.A. 22$^\circ$). We further compute the magnitude
within two apertures (of $r_e$/2 radius) on both sides, east and
west, of the ring (P.A. 94$^\circ$). The magnitudes calculated with
different apertures for \MCG are given in Table~\ref{table4}.

Since NGC~5329 does not show peculiar features in its FUV images and its
emission mainly originates in its central region, we calculated
magnitudes centered on the nucleus within an aperture of 14\arcsec\
diameter, to match the 2MASS measurements and therefore  extend the
data-set from FUV to NIR.

Furthermore, along the tidal tail of NGC~1210 there are knots as
blue as (FUV-NUV)$\approx$0.2, but no obvious counterparts are
visible in the optical frame (see Fig.~\ref{fig3}).  Table~\ref{table5}
presents the magnitudes and colours of the knots, whereas
Fig. ~\ref{fig3} (bottom right panel) displays their position. We
checked, using stellar catalogues, that no known stars contaminate
the knots.

To interpret the properties of  NGC~1210, we adopt the same
photometric tool (stellar models and isochrones,  library of stellar
spectra, and the ABmag photometric system) that has been used to
calculate theoretical {\it GALEX} FUV and NUV magnitudes for SSPs in
\citet{Rampazzo07}.  We are aware that while the Lick line-strength
indices (definited to be the ratio of a central bandpass and two
pseudo-continuum bandpasses on either side of the central band, all
bandpasses being a few tens of \AA) are basically insensitive to
dust attenuation, which could affect the broad-band colours.
The effect on  ages estimated from (FUV-NUV) and (UV-V) colours,  is
discussed in the next paragraphs.

We first derive  the age and  metallicity dependence of the
line-strength  indices and ultraviolet colours and apply it to
NGC~1210. Specifically, we study (i) the relationship between Mg2 and both (FUV-V)
and (NUV-V) colours (see Fig.~\ref{fig8a})  and (ii) the diagnostic
plane H$\beta$ vs  (FUV-NUV) colour, for the particular case of a 10
Gyr old galaxy that experienced a recent burst of a star formation (see
Fig.~\ref{fig8b}).  Looking at the position of NGC~1210 in the two
diagnostic planes of Fig. \ref{fig8a}, the stellar content of this
galaxy is compatible with a young age of about 2-4 Gyr and a mean
metallicity in the range $0.04 < Z < 0.07$. The idea that the whole
stellar content of  NGC~1210 can be as young as 2-4 Gyr is
perhaps too extreme. So it is worth exploring another possibility:
the bulk of the stellar content of the galaxy is old but it suffered a
recent burst of star activity and the age we infer is
essentially that of the young component. This is the motivation for
the experiment shown in the diagnostic plane H$\beta$ versus (FUV-NUV) of
Fig. \ref{fig8b} in which we combine an old population with a young
one and follow the temporal evolution of the composite system.
Examining Fig. \ref{fig8b} in some detail, the position of the 10 Gyr
old model galaxy  is shown by the empty star in the top left corner.
The past evolutionary path of the old object is indicated by a
solid line along which the arrow labelled ``age" is drawn. If a burst
of activity is assumed to occur, the path followed by the model
galaxy is a sort of loop from  the 10 Gyr position back to a new
position very close to it  after a certain time of the order of
about 3 Gyr. The path is clockwise. The various curves correspond to
different choices for the mean metallicity and relative amount of
mass engaged in the novel stellar activity as indicated in the
insert. Once more, the position of NGC~1210 is compatible with a young
age that we now interpret as probably being caused by burst of star formation
of a certain intensity, even if the possibility that the entire 
stellar content is really young cannot be excluded. NGC~1210 indeed lies
close to the fading line of the prototype SSP at an age of
about 4 Gyr.

To conclude,  the positions of the nucleus of NGC~1210 in the planes
Mg2-(FUV-V), Mg2-(NUV-V), and H$\beta$-(FUV-NUV) (Figs. \ref{fig8a} and
\ref{fig8b}) suggest that it hosts a young stellar population, which 
probably formed during the accretion episode. In \citet{Longhetti00}, the value
of H$\beta$ {\it was not corrected} for emission. The correction for
the presence of an emission component within the absorption line
would move the H$\beta$  toward higher values, i.e., towards a
position similar to that of NGC~2865 suggesting the contribution of
a young ($\leq$ 2 Gyr old) stellar population.

\begin{table}
\caption{Magnitudes of the selected circular regions on \MCG}
\begin{tabular}{lccccc}
\hline
band    & HG$_c$         & HG$_N$& HG$_S$& PR$_E$& PR$_W$\\
\hline
FUV & 22.50$\pm$0.15 & 25.32 & 24.41 & 21.55 & 22.40 \\
NUV & 21.07$\pm$0.06 & 23.17 & 22.30 & 21.24 & 21.61 \\
B   & 15.09$\pm$0.02 & 17.64 & 17.57 & 17.56 & 17.75 \\
J   & 13.63$\pm$0.03 & 16.49 & 16.34 & 16.28 & 16.36 \\
H   & 13.30$\pm$0.03 & 16.32 & 16.32 & 16.76 & 17.10 \\
K   & 13.13$\pm$0.02 & 16.18 & 16.25 & 16.77 & 17.06 \\
\hline
\end{tabular}

Note: the radius of the circular regions PR$_E$
(East) and PR$_W$ (West) is r$_e$/2 whereas the radius of HG$_c$ (center), HG$_N$
(North), HG$_S$ (South) on the galaxy body is r$_e$/4,  r$_e$ being
the galaxy effective radius. The typical errors are provided in
col. 1. \label{table4}
\end{table}

\begin{figure}
\includegraphics[width=9.1cm]{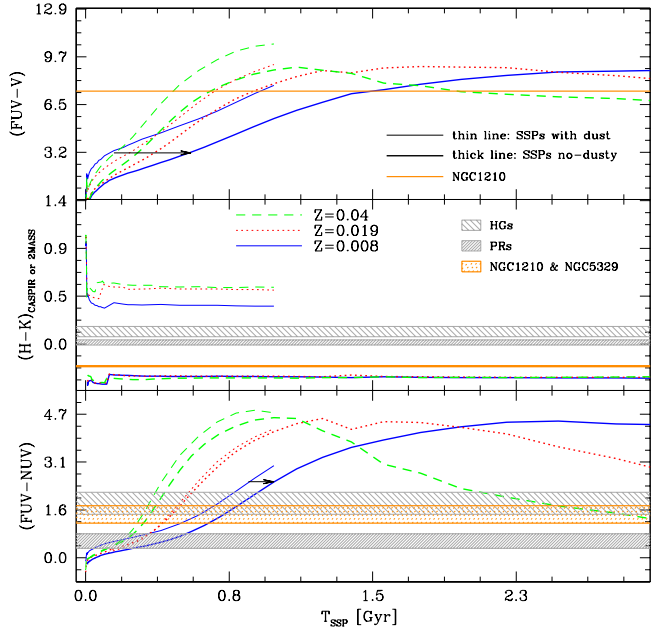}
\caption{The effect of dust on SSPs colours. The shaded bands
represent the colours of NGC~1210, \MCG\ and NGC~5329. \textbf{ Top
panel}: Evolution of the (FUV - V), (H - K) and (FUV - NUV) colours
for ``classical", dust-free SSPs (thick line) and dusty SSPs (thin
line with $\tau_{V}=35$ \citep{Piovan06a}) both shown for  three
different metallicities: Z=0.008, 0.019 (solar), and 0.04 (solid,
dotted, and dashed lines, respectively). \textbf{ Bottom panel}: the
same but for the {\it GALEX} (FUV - NUV) colour. The arrows, in the
top and bottom panels, indicate the ``uncertainty"  in the age
estimate passing  dusty to dust-free  SSPs at given colours.}
\label{fig9}
\end{figure}

\begin{table}
\caption{Magnitudes of the knots along the tail of NGC~1210}
\begin{tabular}{cccc}
\hline
region    &  FUV    & NUV & (FUV-NUV) \\
\hline
a        & 22.64$\pm$0.17 &  21.36$\pm$0.07 & 1.28$\pm$0.18 \\
b        & 22.75$\pm$0.17 &  21.57$\pm$0.07 & 1.18$\pm$0.19 \\
c        & 21.92$\pm$0.15 &  21.58$\pm$0.07 & 0.35$\pm$0.16 \\
d        & 20.92$\pm$0.12 &  20.69$\pm$0.06 & 0.22$\pm$0.13 \\
e        & 21.96$\pm$0.15 &  20.75$\pm$0.06 & 1.21$\pm$0.16 \\
f        & 21.86$\pm$0.14 &  20.18$\pm$0.06 & 1.68$\pm$0.15  \\
g        & 21.44$\pm$0.13 &  19.64$\pm$0.05 & 1.80$\pm$0.14 \\
\hline
\end{tabular}

\medskip
The position of the knots  along the tail in NGC~1210 is shown in
Fig.~\ref{fig3} (bottom right panel). 
\label{table5}
\end{table}

The above estimate of age (and metallicity) of the young stellar
component does not take the presence of dust into account; while the
line-strength indices are not expected to change significantly  in the
presence of dust, the same does not apply to the broad band colours.
Therefore, to properly analyse the effect of the dust attenuation on
the age we need to  use the library of SSP-SEDs  calculated by
\citet{Piovan06a} in which the effects of dust are taken into
account. It is worth recalling here that there are at least three
main circumstances in which dust influences the stellar light:
(i) for a certain fraction of their life very young stars are
embedded in the parental molecular clouds (MCs). Even if the
duration of this obscured period is short, because SNe 
explosions and stellar-wind energy injection  sweep away
the gas soon after the onset of star formation, 
its effect on the light emitted by these young
stars cannot be neglected. Excluding old ellipticals, in which star
formation stopped at relatively early epochs, for all  morphological
types the impact of young dusty populations on the galaxy SED has to
be considered \citep[see][\, for more details about young
SEDs]{Takagi03,Piovan06a,Siebenmorgen07}. (ii) Low- and
intermediate-mass stars in the asymptotic giant branch (AGB) phase
may form an outer dust-rich shell of material obscuring and
reprocessing the radiation emitted by the star underneath
\citep[see][~for the relation between dusty shells around AGB stars
and SSPs]{Bressan98,Piovan03,Marigo08}.(iii) Finally, becauseof the
contribution of metal-rich material by supernovae and stellar winds,
the ISM acquires over the years a dust-rich diffuse component capable
of absorbing/re-emitting radiation \citep[see][\,for models of extinction
and emission of a dusty ISM]{Weingartner01a, Draine01,Li01}.

The \citet{Piovan06a} library of SSPs takes into account the effect
of dust by parental MCs and AGB stars [items (i) and (ii) above]. It
has been successfully employed to reproduce the SEDs of the active
central regions of star-forming galaxies from the UV to the far-IR
by varying  some basic parameters. The parameter space of the
library includes: metallicity $Z$ and age $T_{SSP}$ of the SSP, and
a number of dust-related parameters, namely  the optical depth
$\tau_{V}$ in the $V$-band, the abundance of carbon in the small
carbonaceous grains $b_{C}$ \citep{Weingartner01a}, the scaling
parameter $R$ \citep{Takagi03}, and the ionization state of PAHs
\citep{Weingartner01b}. Since $R$, $b_{C}$, and the ionization state
affect the SED mostly at middle and far-IR wavelengths where we have no data
available for our sample of galaxies, the only parameter of interest is
$\tau_{V}$. This is  chosen to correspond to the case of  highly
obscured, star-forming regions, i.e., $\tau_{V}=35$
\citep{Piovan06a}. The effects of both dust from self-contamination,
parental molecular clouds in which stars are born, and stellar winds
extend from very young ages to ages of about 1 Gyr. Of course,
these SSPs maystill be embedded in a diffuse interstellar component,
which has to be treated separately, but at least for the selfobscuration 
1 Gyr is a safe upper limit to its age \citep[see][ for a detailed
explanation]{Piovan06a, Piovan06b}. With this new library, we  (a)
study the effect of dust extinction on the colours, at least in the age
range 10 Myr to $1$ Gyr and (b) study the effect of a burst of star
formation (by dusty using SSPs) superimposed on an old  dust-free SSP.

\begin{figure}
\includegraphics[width=8.2cm]{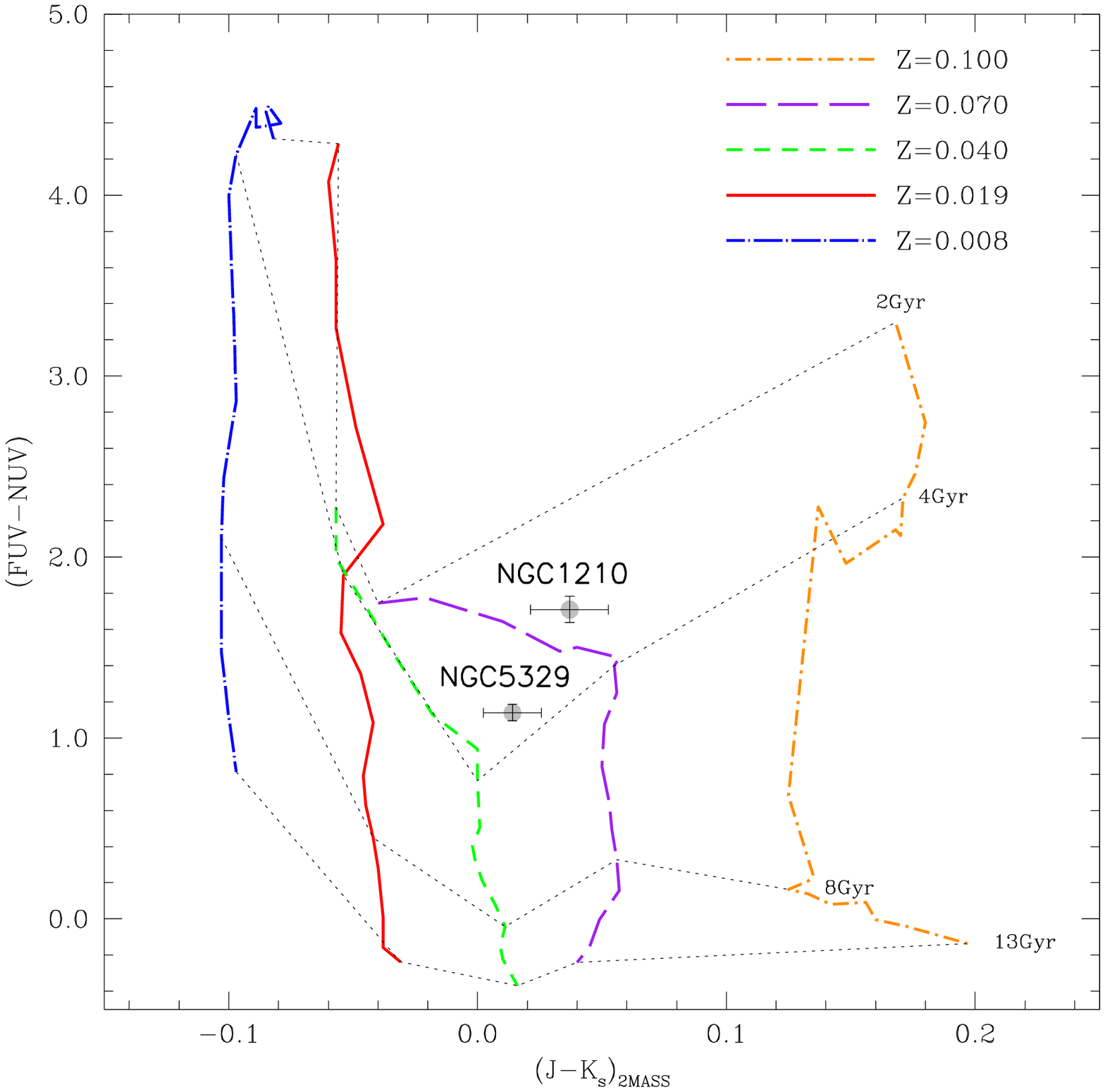}
\caption{Comparison between theoretical colour-colour relations and
{\it GALEX}, optical, and NIR measurements. We plot the {\it GALEX}
(FUV - NUV) vs. (J - K$_s$) theoretical relations for different
metallicities. The loci of constant age are shown for a few selected
age values. No recent bursts are considered. The data for  
the central regions of  NGC~5329 and NGC~1210  suggest that both
galaxies probably experienced  a recent rejuvenation of their stellar
populations.} 
\label{fig10}
\end{figure}

\begin{figure*}
{\centering
\includegraphics[width=9.2truecm,height=7.5truecm]{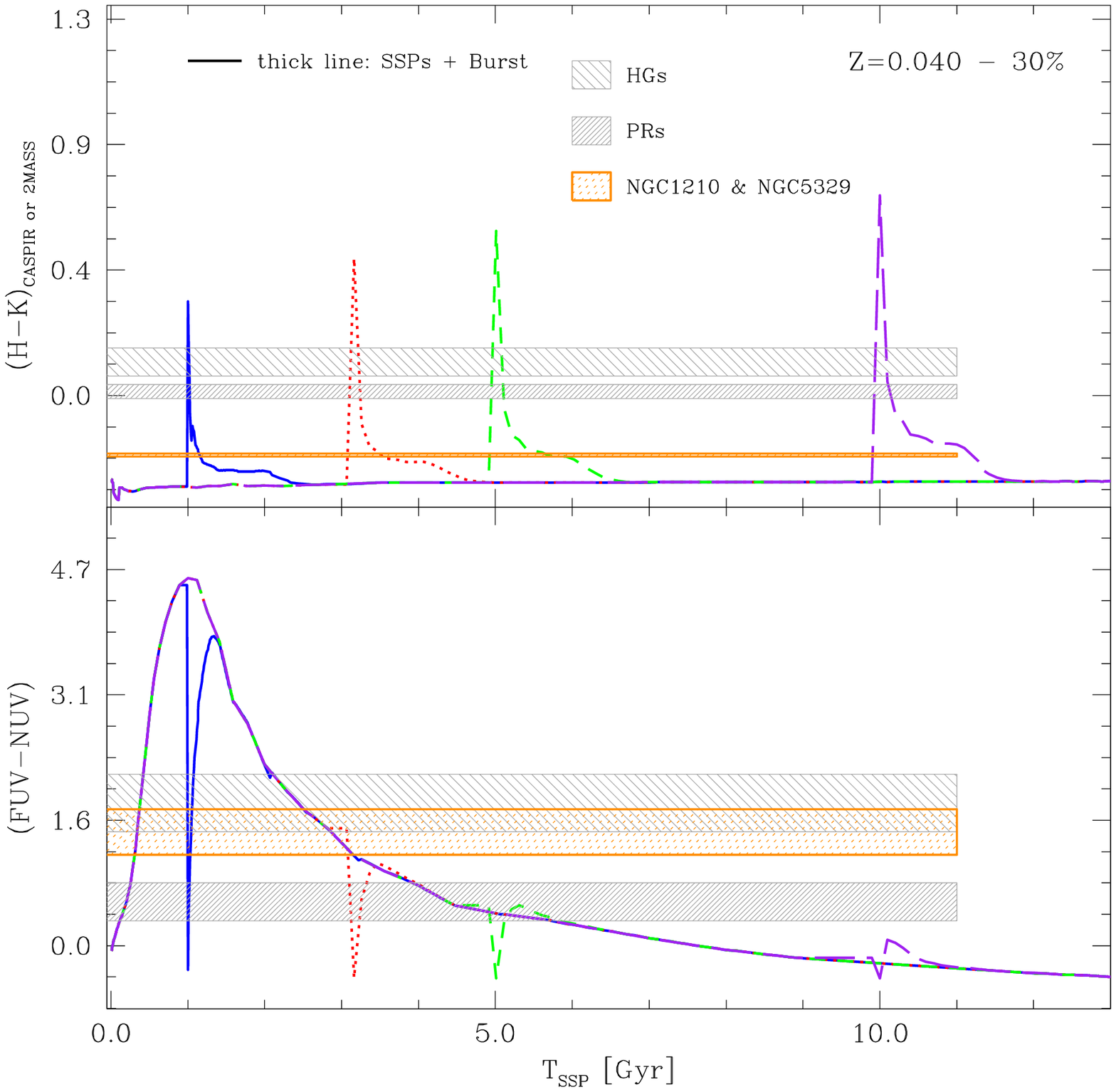}
\includegraphics[width=9.2truecm,height=7.5truecm]{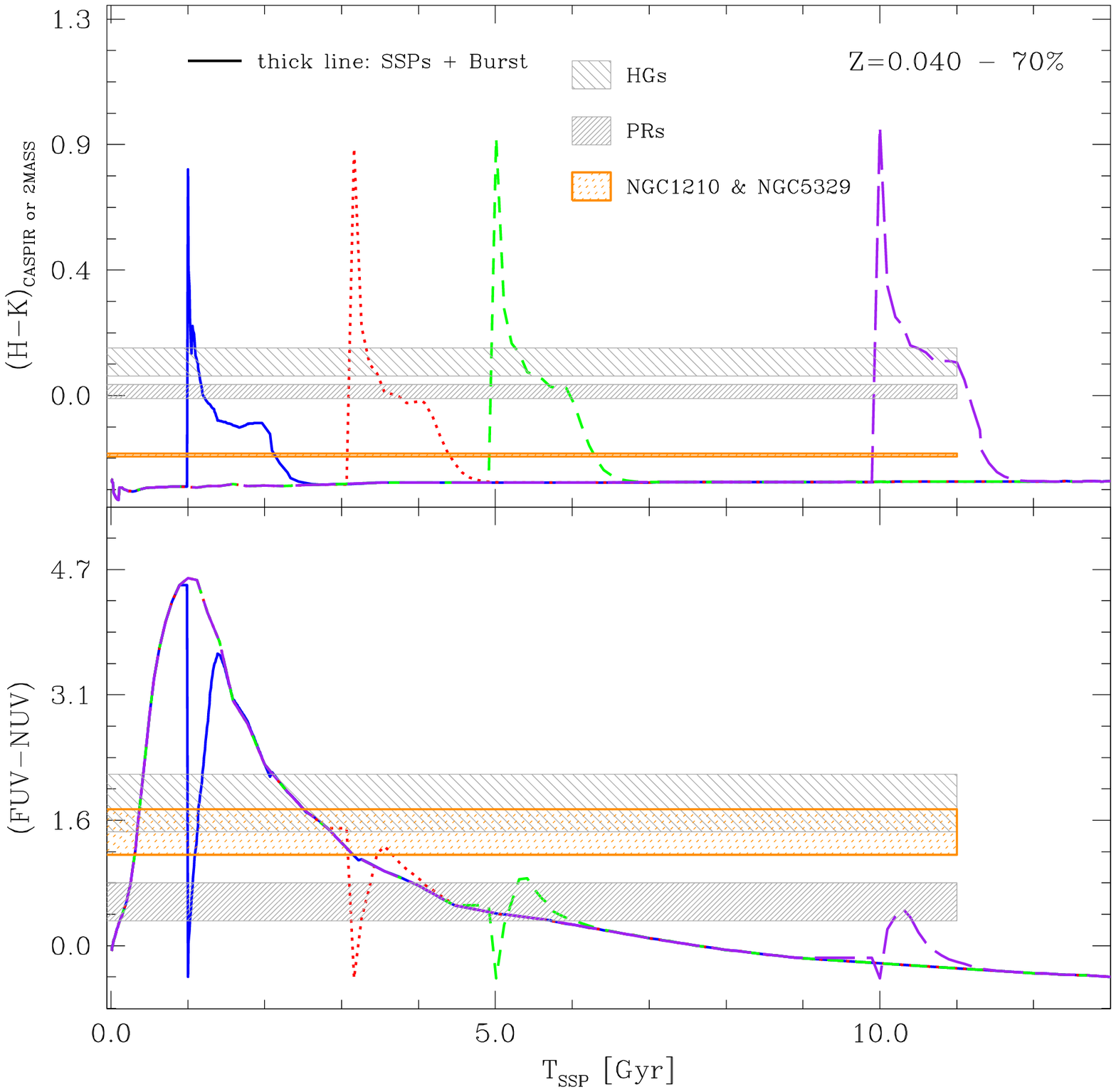}  }
{\centering
\includegraphics[width=9.2truecm,height=7.5truecm]{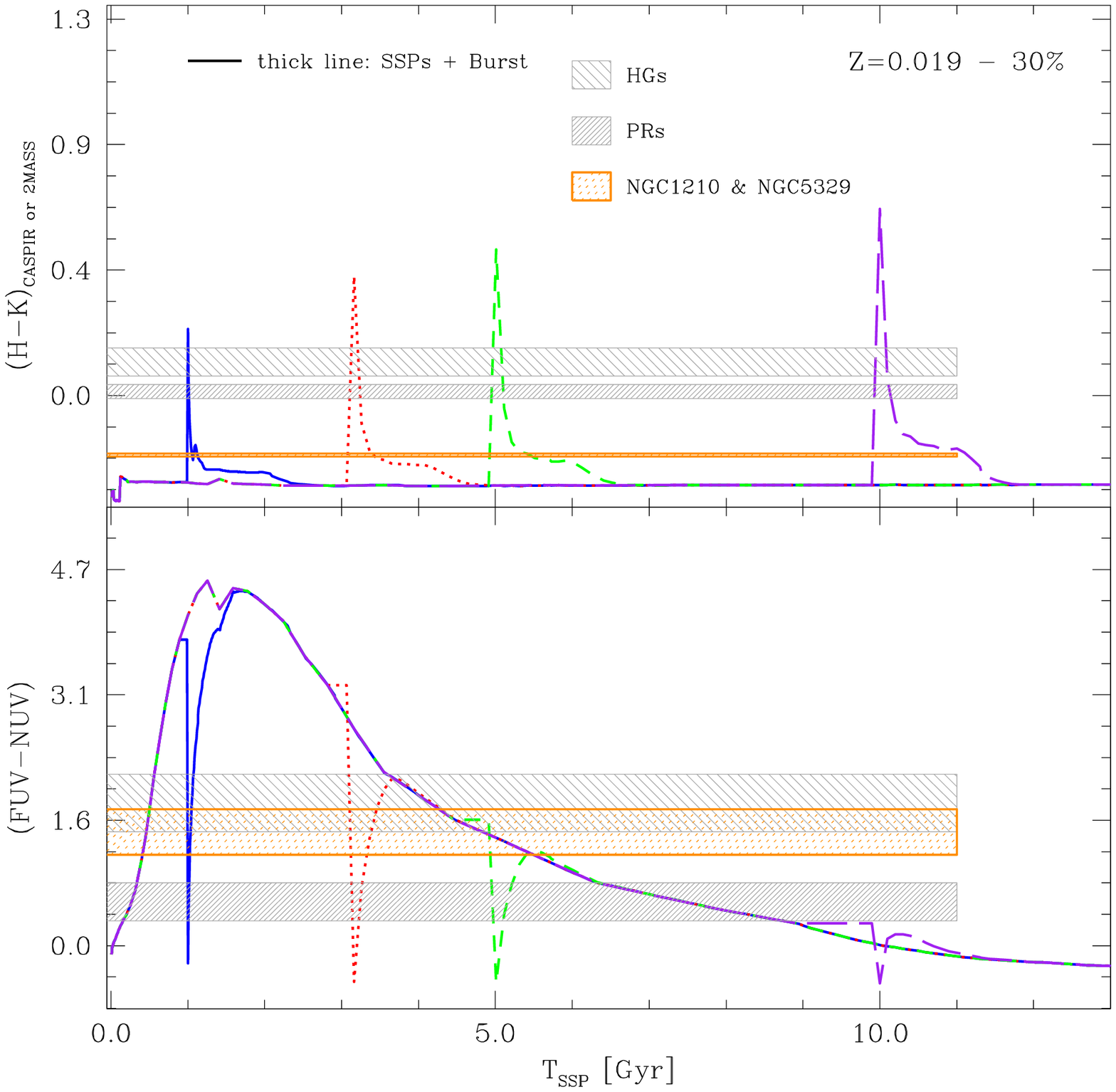}
\includegraphics[width=9.2truecm,height=7.5truecm]{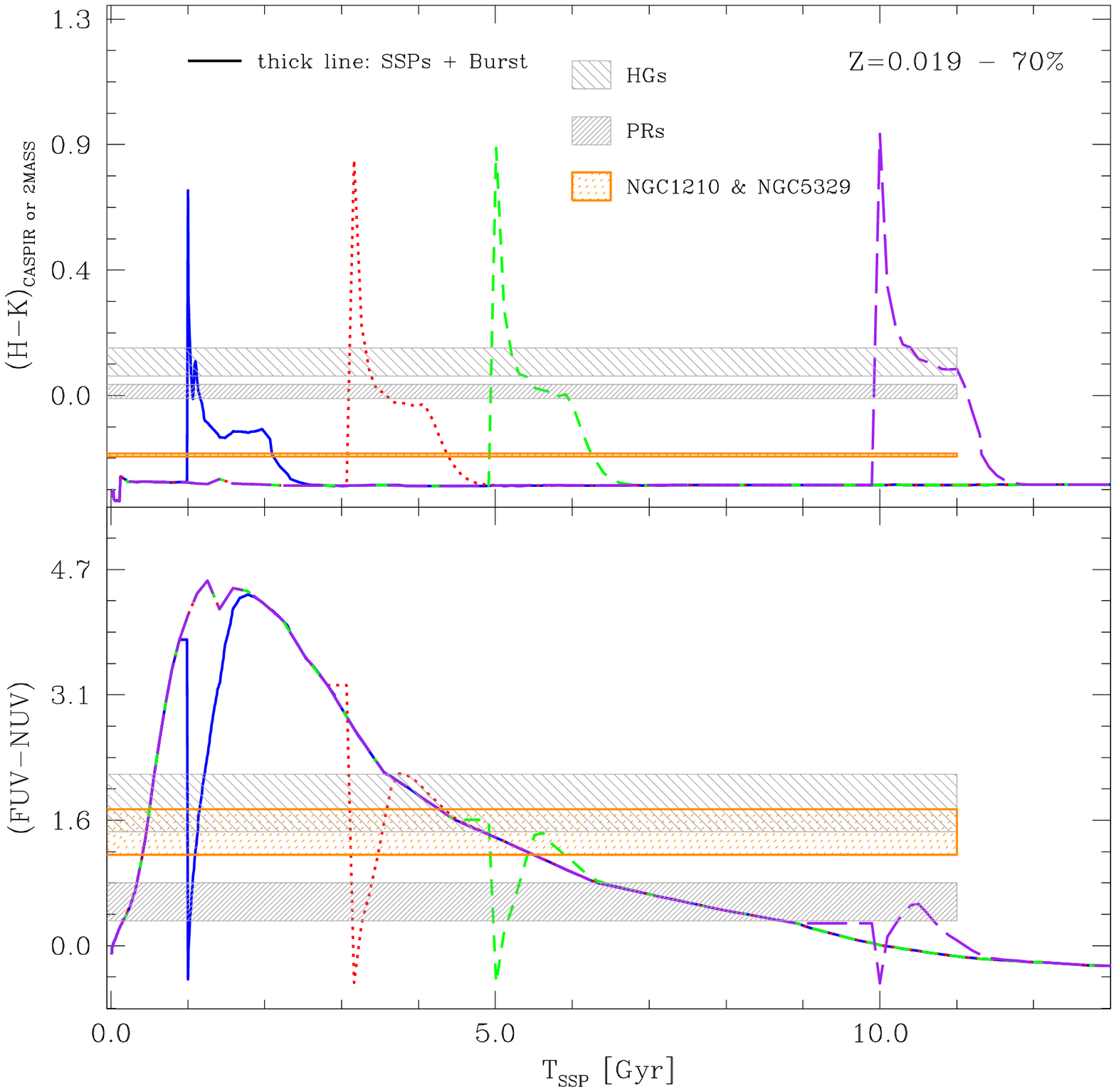}  }
{\centering
\includegraphics[width=9.2truecm,height=7.5truecm]{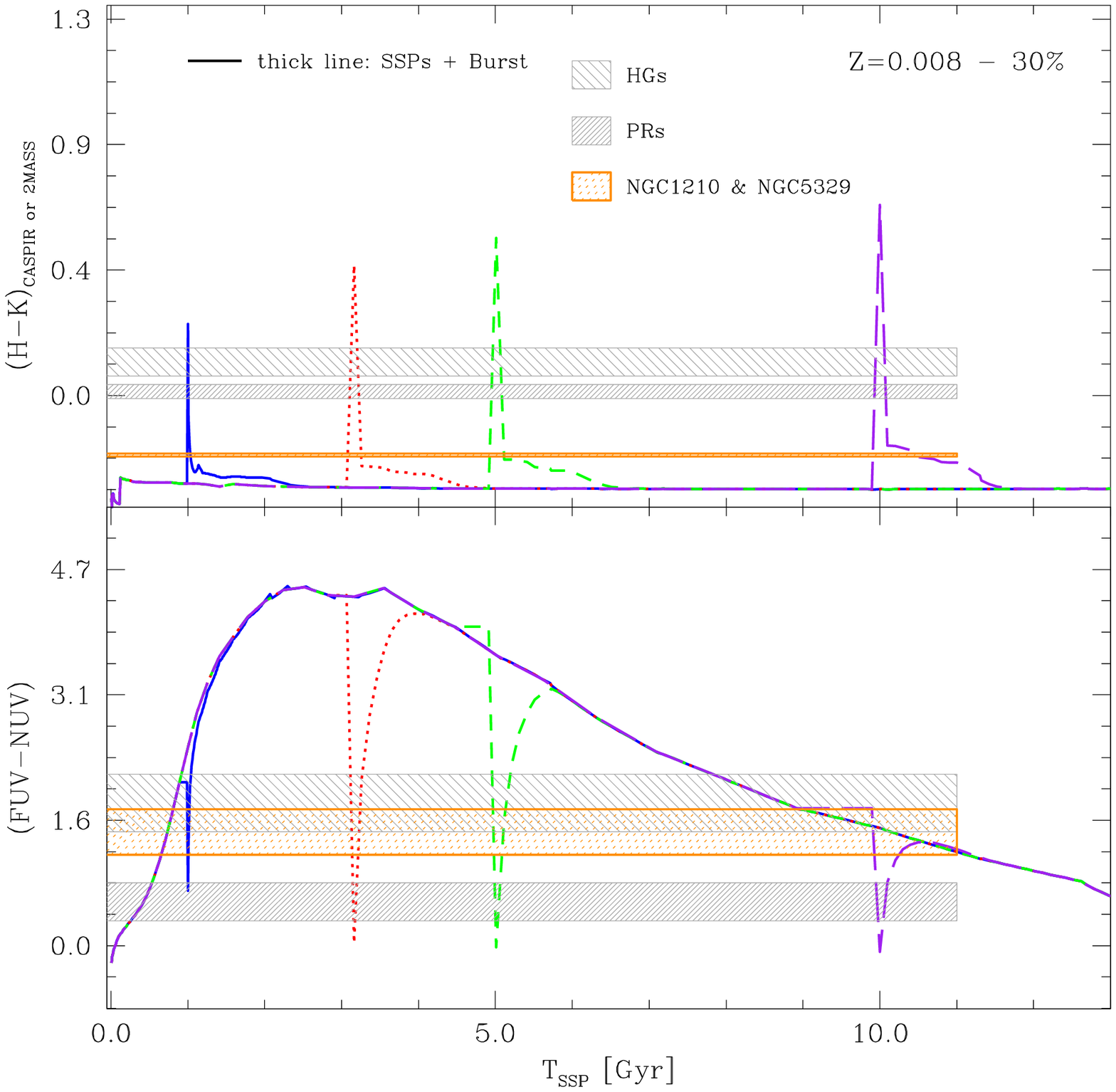}
\includegraphics[width=9.2truecm,height=7.5truecm]{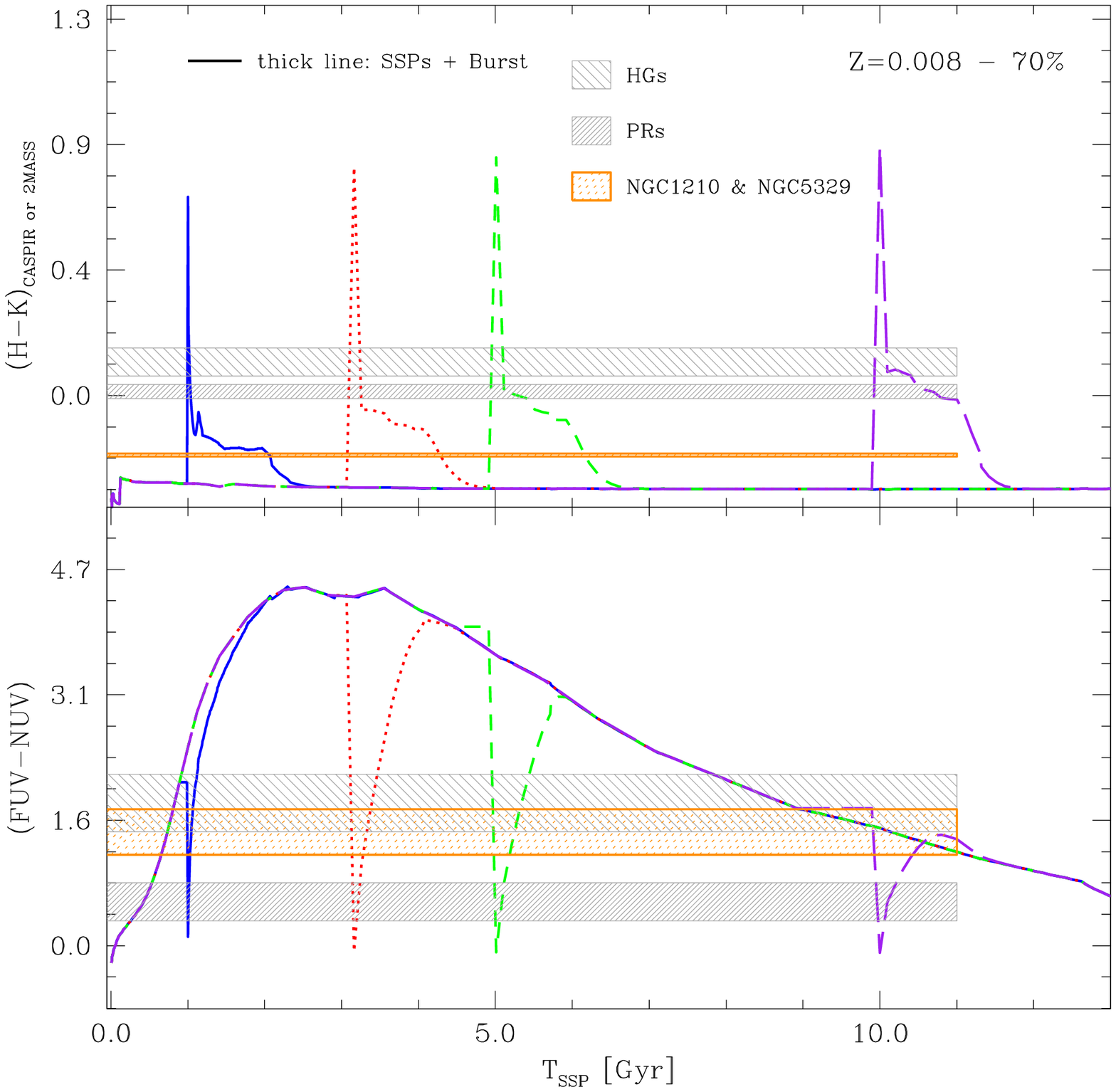}  }
\caption{Evolution of the (H - K) and {\it GALEX} (FUV - NUV) colours
when a burst of star formation is superimposed on the passive
evolution of a classical (dust-free) SSP. The observed colours of
NGC~1210, NGC~5329, and \MCG\ in the host galaxy (HGs) and in the
polar ring (PRs) are shown.
The burst of star formation involves a certain percentage of the 
galaxy mass, namely  30$\%$
({\it left panels}) and of the 70$\%$  ({\it right panels}).  {\it
From, top to the bottom} the burst is superimposed on a 10 Gyr old SSPs
with   metallicities of Z=$0.04$, Z=$0.02$, and Z=$0.008$.} 
\label{fig11}
\end{figure*}

In Fig.~\ref{fig9}, we show the  evolution of the colours (FUV-V),
(FUV-NUV), and (H - K) for both  classical and dusty-SSPs as a function
of different metallicities. The shaded bands indicate the observed
colours of \MCG,  NGC~1210 and NGC~5329.

The attenuation of the UV colours produced by dust introduces an
uncertainty in the age that depends on the metallicity. For (FUV-NUV)
and (FUV-V), the error is as large as $\sim$0.15 Gyr and $\sim$0.5
Gyr for a typical metallicity at Z=0.008 (arrows in the bottom and
top panels of Fig.~\ref{fig9}), respectively. In the mid panel of
Fig.~\ref{fig9}, we show the colour (H-K) for classical and dusty SSPs
as a function of the age and metallicity as indicated. The shaded
band in the same panel show the (H-K) colour measured for NGC~1210,
NGC~5329, and \MCG. To reproduce the NIR colours of the
three sub-regions centered on the northern, central, and southern
regions of the host galaxy \MCG and of the two regions centered on
the western and eastern regions of the ring. Figure~\ref{fig9} shows
that we have to introduce a significant amount of dust in the SSPs.
We conclude that the age estimation of galaxies in the
(FUV-NUV)-H$\beta$ (Fig.~\ref{fig8b}) as well as in the
Mg$_2$-(FUV-V) plane is certainly influenced by  dust attenuation,
the estimated age uncertainty  is any a fraction of a Gyr  (or
smaller given that  we  consider high extinction).

In Fig.~\ref{fig10}, we compare the {\it GALEX} (FUV-NUV) and
(J-K$_n$) colours of NGC~1210 and NGC~5329  with those predicted by
dust-free SSPs. No recent burst of star formation is considered at
this stage of the analysis. The theoretical loci of constant age and
or metallicity are shown as indicated.
 According to mid panel in Fig.~\ref{fig9},  the dust should introduce an
age  uncertainty (small) in the (FUV-NUV) colour and a large effect
in the (J-K)$_{2MASS}$ colour. Since in Fig.~\ref{fig10}, the age follows
a trend nearly perpendicular to (J-K)$_{2MASS}$ axis we suggest that this
should translate into a large uncertainty in the galaxy metallicity
rather than the age. Looking at the colours measured in the
selected regions  centered on the NGC~5329 and NGC~1210 nuclei, we
may conclude that these galaxies probably experienced a recent
($\sim$2-4\,Gyr) rejuvenation of its stellar population. Although
NGC~5329 seems to be  a very relaxed elliptical in the far UV frames, the
estimated luminosity weighted age of the nucleus, inferred from the data,
suggests the presence of a young component in its stellar
population.

For \MCGa,  \citet{Iodice02b}, near-infrared analysis infer
the presence of a very young ($\leq$ 1 Gyr) stellar population (see
their Figs.~2 and 3).

In Fig.~\ref{fig11}, we analyse  the effect of  a secondary burst of
star formation on the colour evolution.  We use 
both dust-free and dusty SSPs to simulate old and recent stellar
components.  We consider  two cases of star-formation bursts.
The first case (left panels) involves 30\% of the stellar population in
the recent burst, and the other (right panel)involves 70 \% of the total in
the burst. Since the bursting activity is accompanied by  dust, the
two cases represent two different situations as far as the total
amount of dust is concerned. Furthermore, the second case can be
considered as an upper limit in which most of the stellar activity
occurs very recently. The shaded areas represent the observed colours
of NGC~1210, NGC~5329, \MCG host galaxy (HGs), and  polar ring
(PRs). The following remarks can be made:

(1) The observed (H - K) infrared colours can be reached by
superimposing a burst of star formation. The NIR colours of both NGC~1210
and NGC~5329 probably can be produced by a weaker burst. The longer the
period of activity or the stronger  the burst, the longer will be  the
period during which the colours match the observed data. Since the
colours of the underlying passive population are almost invariable, the
effect of the burst does not depend  on the burst onset time. Finally,
there are no significant differences between the three
metallicities.

(2) The {\it GALEX} (FUV - NUV) colour during the burst strongly
depends on the metallicity. The higher the metallicity, the earlier
the (FUV - NUV) colour becomes close to the values given by the shaded 
regions of the passive
evolution after a steep initial rise. Therefore, we can set an upper
limit to the age at which the burst began, both for the HG and the
PR regions. For the highest metallicity, the recent episode of star
formation should have began some $<2-3 Gyr$ ago (see the top panel
of  Fig. \ref{fig11}). For  solar and sub-solar metallicities, the
possible age of the secondary stellar activity is older  (see the
mid and bottom panels of  Fig. \ref{fig11}).
With these models, including the effects of  dust, we interpret the
UV, optical, and NIR colours of several selected regions of \MCG 
(see Table \ref{table4}).

\begin{figure}
\centering{
\includegraphics[width=8.4cm]{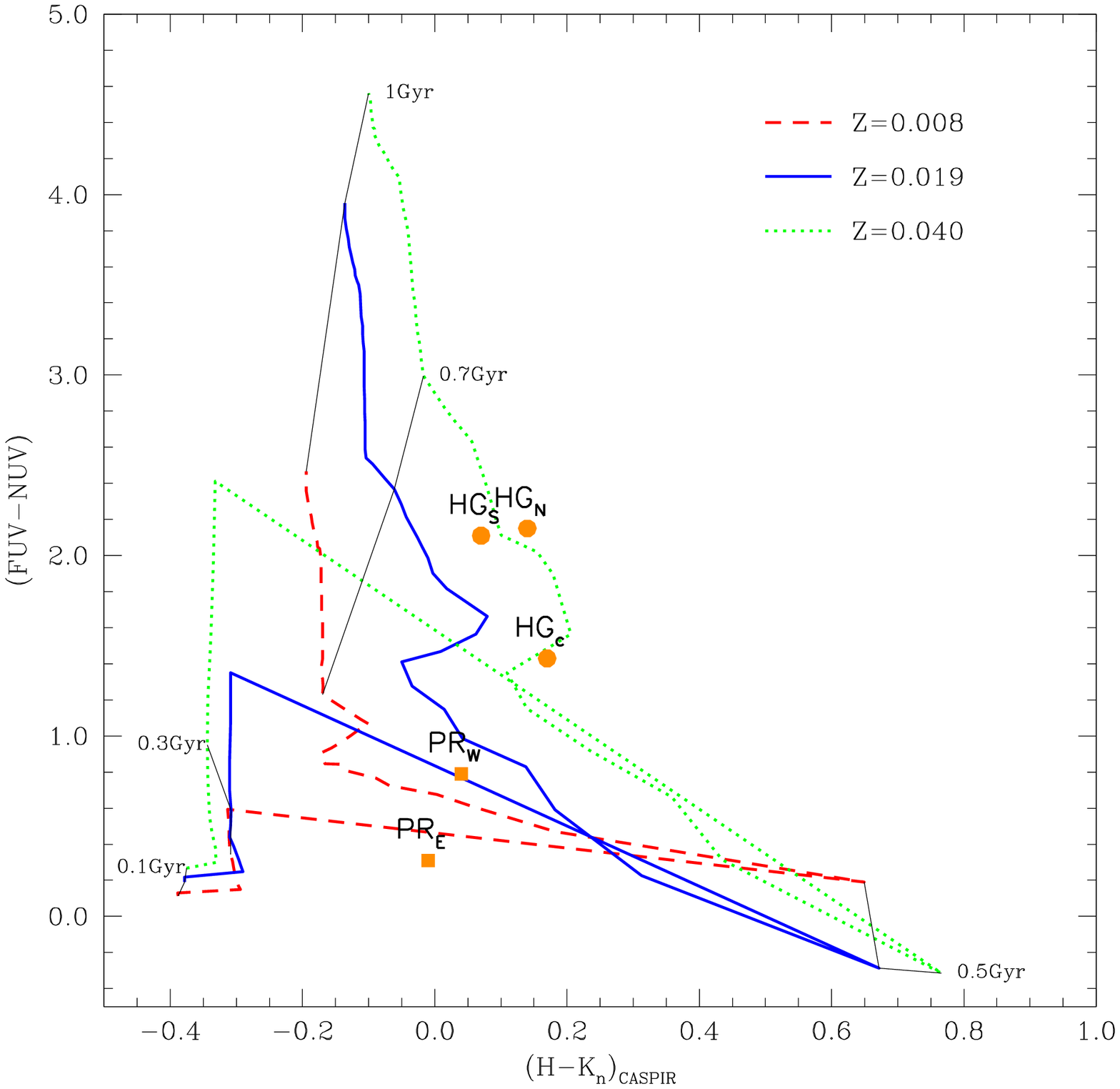}
\includegraphics[width=8.4cm]{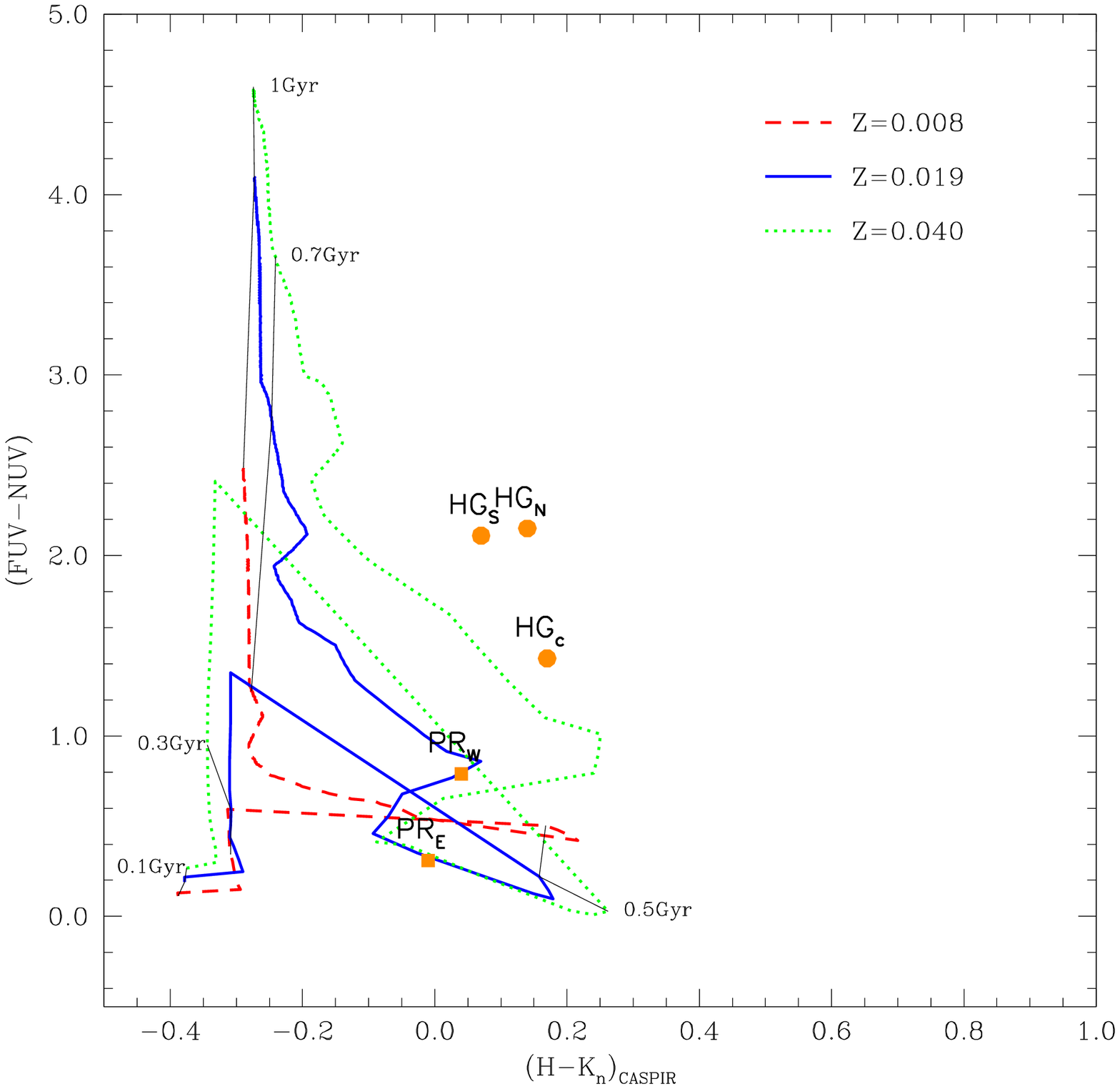}}
\caption{Comparison between theoretical colour-colour relations and
{\it GALEX}, optical, and NIR measurements. We plot the effect of a
recent burst of star formation involving 70\% (upper panel) or 30\%
(lower panel) of the stellar population. SSPs with self-contamination by dust
are considered. The data for the  eastern and western regions of the
\MCGa ring (PR$_{W}$ and PR$_{E}$) and the northern, central, and 
southern regions of the host galaxy (HG$_{N}$, HG$_{C}$, HG$_{S}$)
are shown. The colours of the ring and  host galaxy are indicative the
presence of a very young stellar population.} 
\label{fig12}
\end{figure}

To strengthen the above results, we consider the colour-colour
relationships.  In Fig. \ref{fig12}, we compare the theoretical
two-colour plane {\it GALEX} (FUV - NUV) and (H - K$_{CASPIR}$) with
the observational data   for five selected regions of \MCGa. Two of
them are centered on the eastern and western regions of the \MCG
ring (PR$_{W}$ and PR$_{E}$), whereas the remaining three are
located on the northern, central, and southern regions of the host
galaxy (HG$_{N}$, HG$_{C}$, HG$_{S}$).

We plot the path of a recent burst of star formation (0.5 Gyr old)
of different metallicity $Z$=0.008, 0.019, and 0.04 and different
intensity, namely corresponding to star formation within 30\%  
(bottom panel) and 70\% (top panel) of the
galaxy mass. The SSP in use for the burst are those with dust by
self-contamination. The first choice for the burst intensity is
based on the measured ratio of $H_{2}/\left(H_{2}+HI\right)$ that
could roughly indicate  the amount of gas used up in star formation
if  stars are most likely to form in cold molecular clouds.  With the
second choice, we invert the percentages between dusty and dust-free
SSPs and the composite population  now involves a large fraction,
70\%, of young obscured stars. This latter choice  represents a sort
of upper limit where most of the stars are young and obscured.

As already found   in Fig. \ref{fig11}, there are many
possible ways of combining populations of different age, metallicity,
and total mass. The particular cases shown  in Fig. \ref{fig12}
refer to a metal-rich, relatively young stellar component to which
a burst is added. The colour-colour diagram of Fig.
\ref{fig12} represents an extension in Fig. \ref{fig11}:
different percentages of dusty SSPs are needed to reproduce the
observed colours of PRs and HGs. In general,  a  burst makes  colours
redder  in agreement with the observations. To conclude, we 
 note that even the colour-colour diagnostic  cannot unequivocally determine 
the age  of the underlying stellar population: Figs. \ref{fig11} and
\ref{fig12} clearly show that similar effects are produced  by superposing
a burst on an old metal-poor or a young metal-rich population. This
is the well-known age-metallicity degeneracy, which cannot be broken
in the colour-colour plane {\it GALEX} (FUV - NUV) and (H -
K$_{CASPIR}$). The only conclusion that we can draw is that a strong
contribution by a young obscured population seems to be needed to
reproduce the red colours, but not much can be said about the
underlying population  on which the burst is superimposed.

\subsection{A detailed dusty EPS model for  \MCG }
\label{MCGSED}

The use of both dust-free and dusty SSPs (or SEDs), roughly
mimicking a mixture of old (dust-free) and young (dust embedded)
stellar populations in active galaxies, has provided  
general clues about the age and intensity of the recent burst of
star formation. However, a galaxy is a complex object, whose star
formation history cannot be reduced simply to the sum of two SSPs. For this
reason, we extend the complexity of our analysis by using realistic
evolutionary population synthesis (EPS) models of galaxies and apply
this technique to interpret the data  of \MCGa. For the aims of the
present study, the   EPS  must include the contribution of dust to
the extinction in the UV-optical range. Unfortunately, no data are
available in the MIR/FIR range for \MCGa: these would help us to firmly
constrain the dust content of the galaxy,  the  SED, and finally the
star-formation history (SFH) of this galaxy.

The EPS  models in use here are based on a robust model of chemical
evolution that takes into account both dusty and dust-free SSPs, and
include a realistic description of the geometrical shape of a galaxy
in which stars and dust are distributed \citep{Silva98,Piovan06b}.
The  chemical model includes prescriptions for the infall of
primordial gas, the initial mass function (IMF), the star-formation
rate (SFR), and the stellar ejecta and provides the total amounts of gas
and stars present at any age together with their chemical history
\citep[see][\, for more details]{Tantalo96}. The chemical model
provides the enrichment law $Z(t)$, the SFH, and the gas content to
be supplied to the EPS model (with dust).

The model that we adopt to analyse \MCG\ takes into account the three
different components of the dust: (i) the diffuse interstellar
medium (ISM) composed of gas and dust, (ii) the star-forming regions such
as the large complexes of MCs and, finally, (iii) stars of any age
and chemical composition. The library of dusty SEDs of
\citet{Piovan06a}, described in Sect. \ref{SSPanalysis} is used to
model young star-forming regions, whereas the dust-free SSPs are
included for the stars already free from the parental MCs.

The dust description for the diffuse ISM is included according to
the widely used silicates-graphite-PAH paradigm. The properties of
dust follow \citet{Weingartner01a} for the distribution of the
grains, and \citet{Li01} for the optical properties. To describe the
evolution of the extinction curve with metallicity we use the
sequence of models proposed by \citet{Weingartner01a} that are able to
reproduce the SMC/LMC/MW properties. Indeed, SMC, LMC and MW
represent a sequence of models of increasing metallicity, which
together with the different composition pattern is the origin of
their peculiar average extinction law \citep{Calzetti94}. For the MW
metallicity in particular, we adopted $R_{v}=3.1$ for the diffuse ISM
and $R_{v}=5.5$ for young dusty MCs. Also, eventualy for
Z$>$Z$_{\odot}$, the dust-to-gas ratio scales linearly according
to $\delta=\delta_{\odot}(Z/Z_{\odot}$ but for the same
abundance pattern as the MW. The emission properties of graphite,
silicates, and PAHs are computed according to
\citet{Puget85, Guhathakurta89, Weingartner01b, Draine01,Li01}
\citep[see][\,for a more detailed
explanation]{Piovan06a, Piovan06b}.

The extinction in our model therefore works in two phases: (i) young
stars are attenuated by the  parental MC until the cloud is removed 
by SN explosions and stellar winds from high-mass stars:
(ii) all the stars of any age and metallicity are attenuated by the
diffuse ISM \citep[see][, for other examples of similar attenuation
schemes]{Silva98,Charlot00}.

Finally, to calculate the outgoing radiation we need to describe the
interplay between stars and ISM in producing up the total SED. To
this aim, it is necessary to know how  the stars and ISM are distributed
across a galaxy. The total mass of the gas and stars provided by the
chemical model are then distributed over the entire galaxy volume by
means of suitable density profiles  that depend on the galaxy type
(spheroidal, disc, and disc plus bulge). To calculate the total SED,
the galaxy model is divided into volume elements each of which is,
at the same time, a source of radiation from the inner components
inside and the absorber/emitter of radiation from and to all other
volume elements \citep[see][\, for more
details]{Piovan06a,Piovan06b}.

We adapt the above model to reproduce features of \MCGa, which 
has a total
mass $M_{tot}\sim$1.3$\times 10^{10}M_{\odot}$.   From
Table~\ref{table1}, we know that the total mass of gas ($M_{gas}$) is
$\sim 8\times 10^{9} M_{\odot}$, assuming that
$M_{gas}=M_{HI}+M_{H_{2}}$. The gas and total mass are taken as
basic constraints on the chemical model best suited to represent
\MCG \citep{Tantalo96}, which is imposed to match the ratio
$M_{gas}/M_{tot}\sim 0.6$.

The key properties of this model are listed in the top part of
Table~\ref{table6}, whereas the bottom part lists the key parameters
adopted to simulate the effect of dust \citep[see][\, for all
details]{Piovan06b}. Although \MCG has a complex geometrical
structure, for the purposes of a qualitative analysis, and to distribute
gas (dust) and stars across the galaxy, we can adopt the King profile
\citep{Piovan06b}. Furthermore, the relative proportion  of gas in
the diffuse ISM and  cold MCs given in Table \ref{table1} is
implied by the observational data, as  for the  galaxy
radius.  This is derived from the observational data and set to be equal
to two times the effective radius $r_{e}$, and  is
about 7.8 kpc according to \citet{Whitmore87}. The MC evaporation
time is chosen according to the average lifetime of the most
massive stars, on the notion that this is roughly the timescale
required by stars to expel their parental cloud.

\renewcommand{\arraystretch}{1.3}

\begin{table}
\begin{center}
\caption{Relevant properties of the adopted chemical and dusty
model for the simulated galaxy}
\label{table6}
\begin{tabular}{ll}
\hline\hline
\multicolumn{2}{c}{\textbf{ The basic chemical  model}  }\\
\hline
Total Mass of Galaxy $M_{tot}$                 & $1.30\times 10^{10} M_{\odot}$ \\
Total Mass of Gas    $M_{gas}$                 & $7.82\times 10^{9} M_{\odot}$  \\
Total Mass of Stars  $M_{tot}$                 & $5.06\times 10^{9} M_{\odot}$  \\
Fraction of Gas $M_{gas}/M_{tot}$              & 0.6                            \\
Metallicity $Z$                                & 0.0178                         \\
Time of Galactic Wind$^{a}$                    & 1.09 Gyr                       \\
\hline
\multicolumn{2}{c}{\textbf{ The associated EPS  model with dust} }\\
\hline
Total  Radius                                  &    15 Kpc                      \\
Fraction of Molecular Gas $M_{mol}/M_{gas}$    &    0.3                         \\
Evaporation Time of MCs$^b$                    &    30 Myr                      \\
\hline
\end{tabular}
\begin{minipage}{0.4 \textwidth}
\vspace{1.2mm}
\footnotesize{$^{a}$ Duration of the SFH;\\
$^{b}$ Time scale needed to evaporate the MC in which  a
new generation of stars is embedded.}\\
\end{minipage}
\end{center}
\end{table}

\begin{figure*}
\centering
\includegraphics[width=12.0cm]{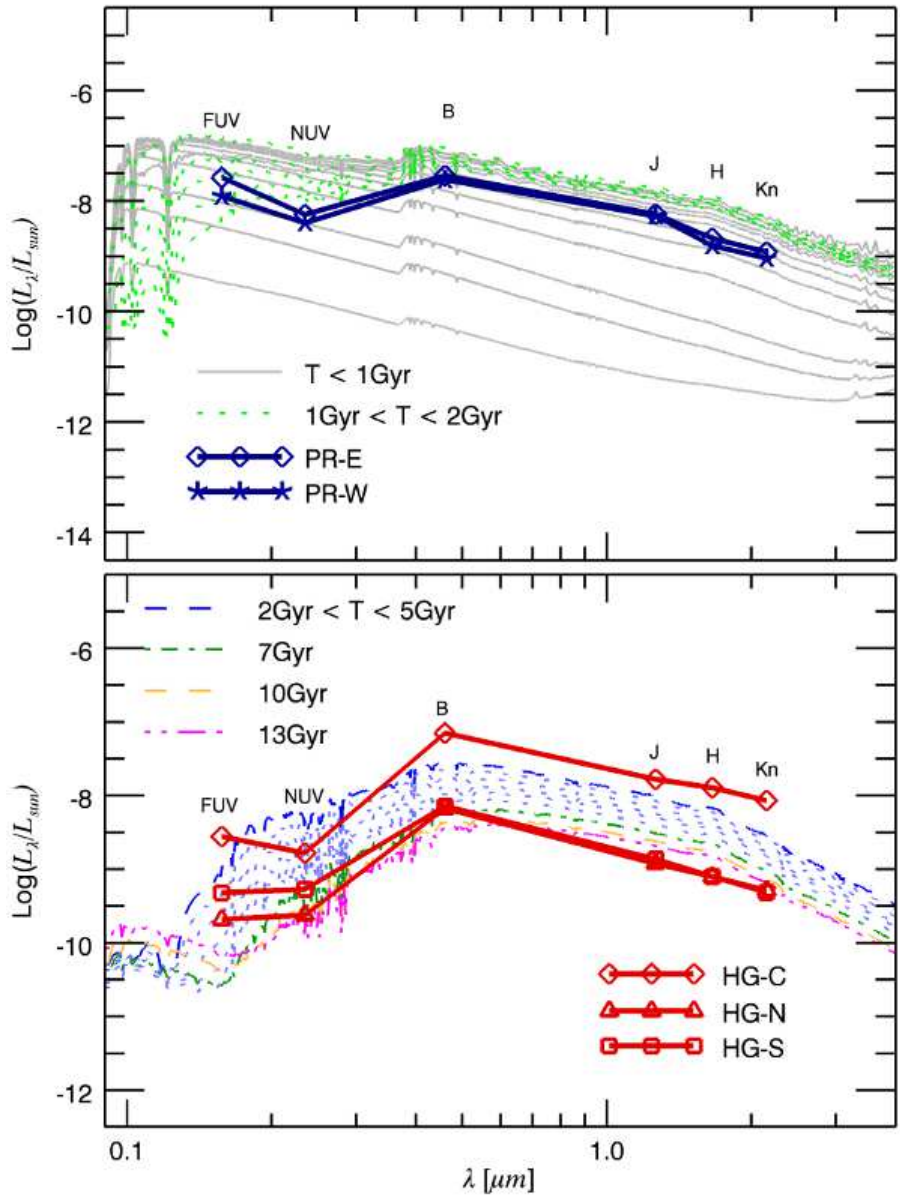}
\caption{Dusty EPS galaxy models of different ages are superimposed on
the  SED of \MCG for which  six observational measurements 
from the far-UV to the near IR are available. \textbf{Top panel}:
Comparison between the theoretical spectra up to the age of 2\,Gyr
and the aperture magnitudes of the east and west sides of the polar
ring. \textbf{Bottom panel}: Comparison between the theoretical
spectra in the range 2 -- 13\, Gyr and the aperture magnitudes
derived on the northern, central and southern regions of the body of
the  host galaxy  along the major axis (see the text for details).}
\label{fig13}
\end{figure*}

\begin{figure}
\centering
\includegraphics[width=8.5cm]{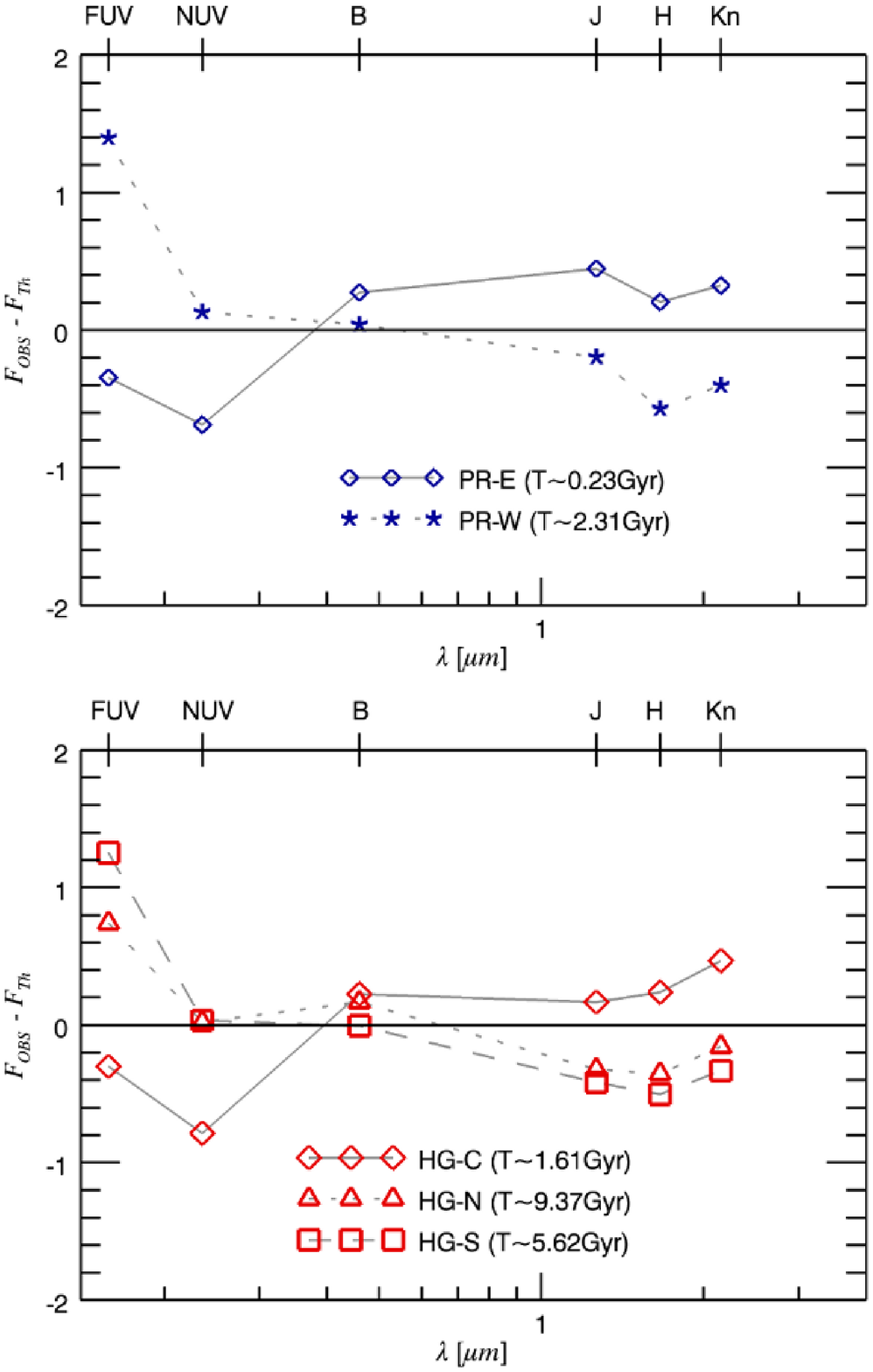}
\caption{Residuals between the dusty EPS galaxy models and
observational data for the six observational data available for
\MCGa, from the far-UV to the near IR. Only the best-fit models
are presented together with the derived age. \textbf{Top panel}:
Residuals between the theoretical SED and the aperture magnitudes of
the eastern (PR-E) and western part (PE-W) of the polar ring.
\textbf{Bottom panel}: Residuals between the theoretical SED and the
aperture magnitudes of  the northern (HG-N), central (HG-C), and
southern (HG-S) regions of the body of the host galaxy  along the
major axis.} 
\label{fig14}
\end{figure}

For the three selected areas across the galaxy for which we have
measurements of the flux (magnitudes) in six different photometric
bands from the far UV to the near IR (FUV, NUV, B, J, H, Kn), we
compare the observational magnitudes with the theoretical
counterparts derived from the chemo-spectro-photometric models of
the galaxy presented above. To this aim, an important quantity to
know is the fraction of galaxy mass encompassed by the aperture in
usage. To estimate the mass, we calculate  how much galaxy light
falls within the aperture located at the two sides of the ring and
the main body of the galaxy, which is approximated to be an oblate disk.
We then calculate the amount of stellar light integrated along the
line-of-sight using the formalism developed by \citet{Galletta84}.
In brief, we assume: 1) an oblate disk with intrinsic axial ratio
$0.25$; 2) an exponential intrinsic density profile, with the same
effective radius of the \MCG\ luminosity profile (19\arcsec); 3) a
constant M/L ratio across the galaxy; 4) a total galaxy extension of
$2.35$\arcmin; 5) that the whole galaxy can be described by a model
made of $100$ shells with same intrinsic flattening. From the
apparent axial ratio of $0.35$, we calculate the inclination of the
line-of-sight with respect to the galaxy plane, which amounts to 
 75.4$^\circ$. The circular apertures are approximated by summing up $13$
line-of-sights, each of which are $1\times 1$ pixel in area ($1.5\times
1.5$ \arcsec in our case). A correction has been added to the
integrated values, to take into account the difference between the
area covered by the 13 pixels and that of a perfect circle. The final
result indicates that the aperture pointed toward  the main body
contains $\sim$ 3\% of its total mass (excluding the
ring), whereas the apertures pointed toward the wings each  include
0.7\% of the total mass. This result is not influenced by the
assumption about the disk flattening, as the galaxy is seen nearly
edge-on. The data plotted in Fig.~\ref{fig13} are scaled by taking into
account these percentages.

As far as the corresponding theoretical magnitudes are concerned, we
estimate the flux falling into each aperture in the following way.
We start  from the definition of ABmag, properly rescale for the
expected flux in the aperture, and finally calculate the luminosity in the
passband to be compared with the observational data. The
theoretical and observational results are compared in
Fig.~\ref{fig13}, where the theoretical luminosities (in magnitude
scale) are shown for different values of the age of the EPS model.

The top panel refers to the data for the two windows that we placed on 
the two sides of the \MCG polar ring (labelled PR in
Table~\ref{table5}). The bottom panel shows the same but for the
windows that we placed on the main body of the host galaxy (labelled HG in
Table~\ref{table5}). In this case, the observational sources are
embedded in the HI gas as shown in Fig.~\ref{fig7}. Finally, in
Fig.~\ref{fig14} we show  the difference between the  theoretical
and the observational luminosities. The legend has the same meaning
as in Fig. ~\ref{fig13}.  The age derived for the best-fits is also
indicated.

The trend of the FUV and NUV fluxes in Fig.~\ref{fig13} for the PR
data suggests the presence of a young stellar population
($\leq$1\,Gyr), whereas the data for the remaining passbands from B
to   K$_n$ seem to indicate the presence of a 2 Gyr population. Using a
best-fit procedure, are  found the ages  to be  $0.23$ Gyr and $2.31$
Gyr in the spectral ranges, FUV and NUV, and from B to K$_n$,
respectively, for both PR data. On the other hand, the three SEDs
obtained with apertures located on the HG are more accurately described by
EPS models with  ages in the range 2-9\,Gyr. The best-fit to the
three HG regions are consistent with this age interval, as  shown in
Fig.~\ref{fig14}.

Despite all of these  uncertainties, these results are consistent with those
obtained in Sect. \ref{SSPanalysis} using simple SSPs: the \MCG\
colours can be reproduced by assuming  a recent, active burst of star
formation occurred in the PR superimposed on an older population of stars in
the HG. The analysis carried out with SEDs seems to favour this
possibility, already proposed in Sect.~\ref{SSPanalysis}, with
respect to the one presented in Fig.~\ref{fig12}, where both the
burst and the underlying population were young. Our analysis therefore
suggests that a very recent burst of star formation has
been triggered probably by a gas rich wet accretion event onto an
underlying older stellar system.

Finally, it is worth noting that  even the best-fit to the SEDs
does not give fully  satisfactory results for all  passbands and
that the uncertainties are still large. We tried to vary some basic
parameters of the model, such as the evaporation timescale of
parental MCs,  without substantially improving the results. 
The adopted geometry is probably too simple for a very irregular and complex
system such as \MCGa, for which a proper treatment  would require a more 
complicated description of the spatial distribution of dust and stellar sources.

\section{Summary and conclusions}\label{sum_concl}

Systems of shells and polar rings in early-type galaxies
are considered to be tracers of accretion and merger events.
Their high frequency in low density environments suggests that these
episodes could drive the secular evolution of at least a
fraction of the early-type galaxy population.

We have obtained the surface photometry of  three early-type galaxies
observed with  {\it GALEX}: \MCGa, a well-known
polar-ring galaxy, and NGC~1210 and NGC~5329, which both display 
shell systems.  In the present study, only NGC~5329 belongs to a poor
galaxy structure (WBL~472), the others being very isolated galaxies.
NGC~1210 and \MCG have a large quantity  of HI gas associated
with their stellar body.

Both the {\it GALEX} NUV and FUV images of \MCG\ and NGC~1210 detect
complex tidal tails and debris structures. In the FUV,   
\MCG and NGC~1210 may be hardly classified as ``classical" early-type galaxies from
their morphology. In NGC~1210, inner shells are visible in the NUV
image. The inner shell (labelled with ``2" in Fig.~\ref{fig3}) 
probably has a different origin from the outer ones, as suggested by the
tail-like structure that it has in the FUV image. A case of multiple
accretion has also  been suggested for the shell galaxy NGC~474 by
\citet{Rampazzo06} and  \citet{Sikkema07}.

The polar ring of \MCG is clearly visible in data for both {\it GALEX} bands,
while the host galaxy, from which the bulk of the emission in
optical and near infrared originates is almost undetected.   {\it GALEX} data
of NGC~ 5329 show that the distribution/morphology  of its NUV emission is
comparable to that of the optical image, while FUV emission
is evident only in the central regions of the galaxy. This implies
that the NUV emission originates in the same kind of stellar population.
The FUV emission, which is more concentrated toward the nucleus, suggests that 
different type of hot stars contribute to the flux \citep[see also][]{Rampazzo07}.
NGC~5329 does not show evidence of shells in the {\it GALEX} bands.

We used optical and near-infrared images to build  coherent pictures
of stellar content  in both the nuclear region and  the outskirts of
the galaxies as well as the region where fine structures are
present. We investigate the capability of the combined  far
UV, optical, and NIR colours to provide information about the time at
which the accretion/merging phenomenon  occurred. Our models
suggest that a very young ($\leq$ 1 Gyr) stellar population is present
in \MCGa: we cannot exclude the possibility of ongoing star formation in
this galaxy.  The nuclei of NGC~1210 and NGC~5329 also appear
rejuvenated but the accretion episode that triggered these star
formation episodes should be much older (2-4 Gyrs).

Furthermore,  the age estimated for the polar ring is  
consistent with the stability timescales of the polar structure
formed by the accretion scenario: a low inclined polar ring (i.e.,
nearly polar) may be observed after several Gyr \citep{Bournaud03}.

\citet{vanDokkum05} suggested that dry mergers, i.e. nearly
dissipation-less or gas less mergers at low redshift, are
responsible for much of the local bright field elliptical galaxy
population. Oddly enough, about 50\% of nearby early-type galaxies
were found to contain detectable HI \citep[see
e.g.,][]{Schiminovich01, Serra07a}. \MCG\ and NGC~1210 are examples
of wet mergers: they are dynamically young objects according to their 
irregular HI gas spatial distribution and  young
stellar population.  {\it GALEX} images display
unambiguously the strong morphological connection between the far UV
and the HI distribution shedding light on the secular evolution of
early-type galaxies and the final fate of the accreted gas. The
presence of a young stellar population is then  connected to the
recent accretion and the refuelling of fresh gas to an otherwise
old galaxy, rejuvenating its star formation.

\begin{acknowledgements}
We acknowledge partial financial support of the Agenzia Spaziale
Italiana under contract ASI-INAF contract
I/023/05/0.  A.M. acknowledges partial financial support from 
The Italian Scientists and 
Scholars of North America Foundation (ISSNAF) and J. Herald for help in 
revising the text.  
{\it GALEX} is a NASA Small Explorer, launched in April
2003. {\it GALEX} is operated for NASA by California Institute of
Technology under NASA contract  NAS-98034. This research has made
use of the SAOImage DS9, developed by Smithsonian Astrophysical
Observatory and of the NASA/IPAC Extragalactic Database (NED) which
is operated by the Jet Propulsion Laboratory, California Institute of
Technology, under contract with the National Aeronautics and Space
Administration. {\tt IRAF} is distributed
by the National Optical Astronomy Observatories, which are operated
by the Association of Universities for Research in Astronomy, Inc.,
under cooperative agreement with the National Science Foundation.
We acknowledge the usage of the HyperLeda database (http://leda.univ-lyon1.fr).
The Digitized Sky Survey (DSS) was produced at the
Space Telescope Science Institute under U.S. Government grant NAG
W-2166. The images of these surveys are based on photographic data
obtained using the Oschin Schmidt Telescope at the Palomar
Observatory and the UK Schmidt Telescope. The plates were processed
into the present compressed digital form with the permission of
these institutions.
\end{acknowledgements}

\bibliographystyle{aa}
\bibliography{11819}

\end{document}